\documentclass[review]{elsarticle}

\usepackage{latexsym,graphicx,amssymb,amsmath,mathrsfs}
\usepackage{setspace,bm}
\usepackage{amsmath}
\usepackage[usenames]{color}
\usepackage{latexsym}
\usepackage{epstopdf}
\usepackage{mathtools}
\usepackage{multirow}
\usepackage[breaklinks, colorlinks=true, pdfstartview=FitV, linkcolor=red, citecolor=blue, urlcolor=blue]{hyperref}

\usepackage{lineno}

\modulolinenumbers[5]

\journal{Journal of \LaTeX\ Templates}

\newcommand\beq{\begin{equation}}
\newcommand\eeq{\end{equation}}
\newcommand\beqa{\begin{eqnarray}}
\newcommand\eeqa{\end{eqnarray}}

\usepackage[normalem]{ulem}

\begin{document}

\begin{frontmatter}

\title{Persistent homology analysis of deconfinement transition in effective Polyakov-line model}

 \author[Kyushu]{Takehiro Hirakida\corref{corresponding}}
 \cortext[corresponding]{Corresponding author}
 \ead{hirakida@phys.kyushu-u.ac.jp}

 \author[FIT]{Kouji Kashiwa}

 \author[Kyushu]{Junpei Sugano}

 \author[JMA]{Junichi Takahashi}
 \author[Saga]{Hiroaki Kouno}

 \author[Kyushu]{Masanobu Yahiro}

 \address[Kyushu]{Department of Physics, Graduate School of Sciences, Kyushu University, Fukuoka 819-0395, Japan}
 \address[FIT]{Fukuoka Institute of Technology, Wajiro, Fukuoka 811-0295, Japan}
 \address[JMA]{Numerical Prediction Division, Forecast Department, Japan Meteorological Agency, Tokyo, 100-8122, Japan}
 \address[Saga]{Department of Physics, Saga University, Saga 840-8502, Japan}


\begin{abstract}
 The persistent homology analysis is applied to the effective
 Polyakov-line model on a rectangular lattice to investigate the
 confinement-deconfinement nature.  The lattice data are mapped onto the
 complex Polyakov-line plane without taking the spatial average and then
 the plane is divided into three domains.  This study is based on
 previous studies for the clusters and the percolation properties in
 lattice QCD, but the mathematical method of the analyses are different.
 The spatial distribution of the data in the individual domain is
 analyzed by using the persistent homology to obtain information of the
 multiscale structure of center clusters.  In the confined phase, the
 data in the three domains show the same topological tendency
 characterized by the birth and death times of the holes which are
 estimated via the filtration of the alpha complexes in the data space,
 but do not in the deconfined phase.  By considering the configuration
 averaged ratio of the birth and death times of holes, we can construct
 the nonlocal order-parameter of the confinement-deconfinement
 transition from the multiscale topological properties of center
 clusters.
\end{abstract}


\end{frontmatter}

\section{Introduction}

Topological properties of the system can play a crucial role in the
classification of phase transitions.  There are several ways to apply
topological knowledge in mathematics to physics.  Recently, the
persistent
homology~\cite{edelsbrunner2000topological,zomorodian2005computing},
which is one of the ways to introduce the topological viewpoint to
physics, attracted much more attention in the classification of system's
structure~\cite{nakamura2015persistent,hiraoka2016hierarchical,donato2016persistent}.
In this article, we utilize the persistent homology to investigate phase
structures in the effective Polyakov-line model which is the effective
model of quantum chromodynamics (QCD) in the heavy quark mass regime.

The property of the confinement-deconfinement nature in the pure
Yang-Mills theory and also QCD is a long-standing problem and thus
several proposal have been stated so far; see
Ref.~\cite{greensite2010introduction}.  At least in the pure Yang-Mills
theory, the Polyakov-line (loop) which relates the gauge invariant
holonomy can exactly describe the confinement-deconfinement transition
because it can be expressed by using the one-quark excitation
free-energy.  The finite value of the Polyakov-line also indicates the
spontaneous breaking of $\mathbb{Z}_3$ symmetry.  Thus, we can regard
the Polyakov-line as the order-parameter of the
confinement-deconfinement transition.  However, in QCD, the
Polyakov-line is not an exact order parameter anymore, since the
existence of dynamical quark breaks $\mathbb{Z}_3$ symmetry explicitly.
Hence our understanding of the confinement-deconfinement nature is still
limited.

From the topological viewpoint, it has been recently suggested that the
confinement and deconfinement states of QCD at zero temperature can be
clarified via the topological order~\cite{Sato:2007xc} which is
characterized by the ground-state degeneracy in the compactified
space~\cite{Wen:1989iv}.  Note that the notion of the topological order
is valid even if there is no symmetry that governs the transition.
Analogy of the topological order in QCD has been applied to finite
temperature in
Refs.~\cite{Kashiwa:2015tna,Kashiwa:2016vrl,Kashiwa:2017yvy} by
considering the non-trivial free-energy degeneracy at finite imaginary
chemical potential.  Therefore, it is natural to expect that the
topology can bring us to deeper understanding of the
confinement-deconfinement nature.


Another interesting study based on the topological view point is the
study of the center clusters and the
percolation~\cite{Gattringer:2010ms,Borsanyi:2010cw,Endrodi:2014yaa}.
The center clusters can be classified from the behavior of
Polyakov-line.  In the complex Polyakov-line plane, the spatial
distribution of local Polyakov-line can have the wide spread; see Fig.~2
in Ref.~\cite{Endrodi:2014yaa}.  The distribution of the center cluster
depends on the temperature and have important information of
confinement-deconfinement nature.  It was shown that the fractal
dimension related to the scale properties of the center cluster
distribution is useful to analyze the confinement-deconfinement
transition~\cite{Endrodi:2014yaa}.  It was found that the fractal
dimension changes rapidly through the transition.  Note that the study
has been extended to the case of the full QCD~\cite{Schafer:2015wja}
where there is no exact symmetry that governs the
confinement-deconfinement transition.  Hence, it is naturally expected
that the analyses of the multiscale properties of the center cluster
distribution is important and useful to investigate the
confinement-deconfinement nature of QCD.  Combining the topological
notion and the multiscale properties, in this study, we investigate the
multiscale topological properties of spatial distribution of the center
cluster by another method, namely, the persistent homology to obtain the
deeper understanding of confinement-deconfinement nature.

The persistent homology is widely used to investigate the multiscale
topological structure of the distribution of data set; for example, the
spatial structures of matter such as the glass and liquid states of
SiO$_2$ can be well classified from the persistent homology, even if
these states have almost the same spatial
distributions~\cite{hiraoka2016hierarchical}.  Therefore, it seems to be
suitable to apply the persistent homology analysis to investigate the
structure of the center clusters.

\begin{figure}[h]
 \centering
 \includegraphics[width=0.7\textwidth]{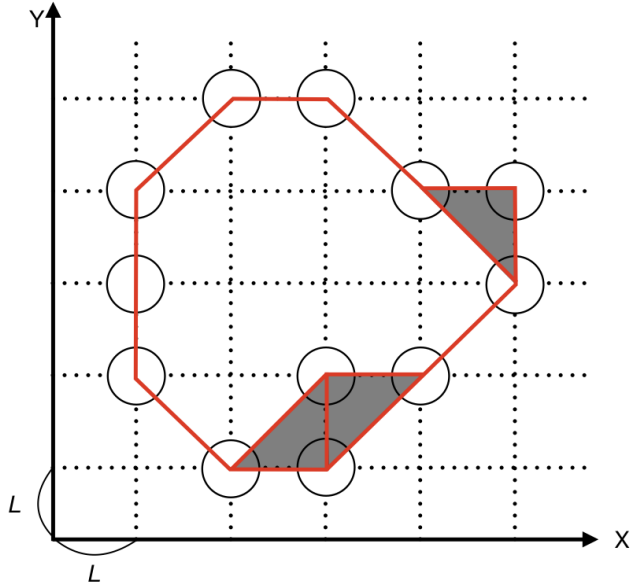}
 \caption{A schematic figure of the multiscale topological structure
 that the persistent homology can extract in the two-dimensional data
 space.
 When the resolution scale is larger than $\sqrt{2}L$ (but
 smaller than the ring scale), we recognize only a ring as a topological
 structure.  } \label{fig:demo_cluster}
\end{figure}

In Fig.~\ref{fig:demo_cluster}, we show a schematic figure of the
multiscale topological structure that the persistent homology can
extract.  In the actual calculation of the persistent homology, we just
need the data set and analyze the data from filtration, as explained
later.  In the calculation, we can prepare several different data forms
such as the bare data, the averaged data, and the mapped data to analyze
the topological structure of data space.  Therefore, we may
systematically analyze the topological structure from the persistent
homology with different data forms.

In this article, as a first step to apply the persistent homology to
QCD, we consider the effective Polyakov-line model as a QCD effective
model in the heavy quark mass regime.  This article is organized as
follows.  In Sec.~\ref{Sec:PH} the persistent homology is explained.
After giving an introductory summary of persistent homology
in Sec.~\ref{SubSec:PH1},
some technical and mathematical explanations are given
in Sec.~\ref{SubSec:PH2}.
The formulation of the effective Polyakov-line model is shown in
Sec.~\ref{Sec:EPL}.  Numerical results are shown in
Sec.~\ref{Sec:Results}.  Section~\ref{Sec:Summary} is devoted to summary.

\section{Persistent homology}
\label{Sec:PH}

\subsection{Introductory summary of persistent homology}
\label{SubSec:PH1}

\begin{figure}[h]
\centering
 \includegraphics[width=0.9\textwidth]{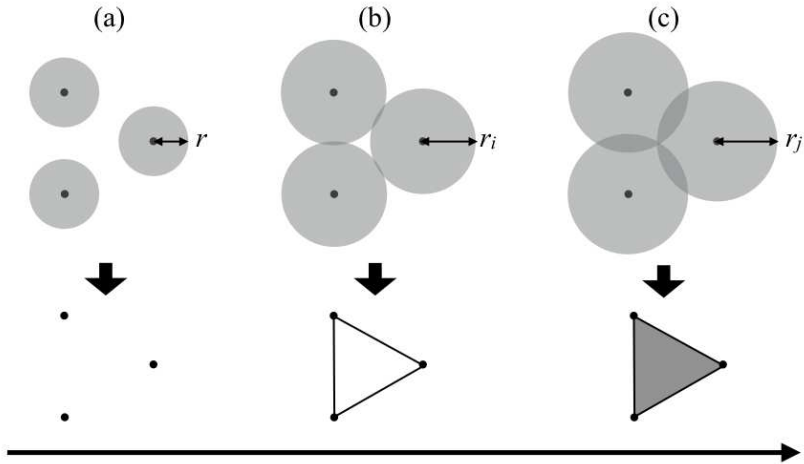}
 \caption{Ball model of three data points, where $r$ denotes the radius of the ball.
Corresponding complexes are also shown.
(a) Three points. (b) Three points and three line segments. (c) Three points, three line segments and a triangle. }
 \label{Filtration}
\end{figure}

First, we show an introduction
of persistent homology.  Technical
and Mathematical details will be given in the next subsection.
We consider the three data points in two dimensional data space, and consider a two
dimensional ``ball'' the center of which is the data point and the
radius of which is denoted by $r$.  As shown in Fig.~\ref{Filtration}
(a), when $r$ is small enough, three data points are isolated each
other.  In this case, we regard the ball as a zero-dimensional point
which is called a ``zero-dimensional simplex'' in the contexture of
homology.

If the radius $r$
increases and becomes larger than the half
of the distance between two data points, two ball overlap as shown in
Fig.~\ref{Filtration} (b).  In this case, we regard that the two data
points are ``connected'' and a line segment newly appears in addition to the
data points themselves.  The line segment is a one-dimensional object and
called a ``one-dimensional simplex''.  Note that, in Fig.~\ref{Filtration} (b), there is
a ``hole'' surrounded by three line segments.  If $r$ becomes larger
than the half of the maximum of distances between two points, three
balls overlap and hole disappears as is shown in Fig.~\ref{Filtration}
(c).  On the other hand, a ``two-dimensional simplex'', namely, a
triangle, newly appears.

As $r$ increases, the number of the simplexes in the diagram increases.
In Fig.~\ref{Filtration} (a) there are only three data points.  In
Fig.~\ref{Filtration} (b), in addition to data points, three line
segments appear.  There are three data points, three line segments and
one triangle in Fig.~\ref{Filtration} (c).  The set of simplexes is
called a ``complex''.  (For rigorous definitions of complex, see the
next subsection.)

\begin{figure}[h]
\centering
 \includegraphics[height=0.5\textheight]{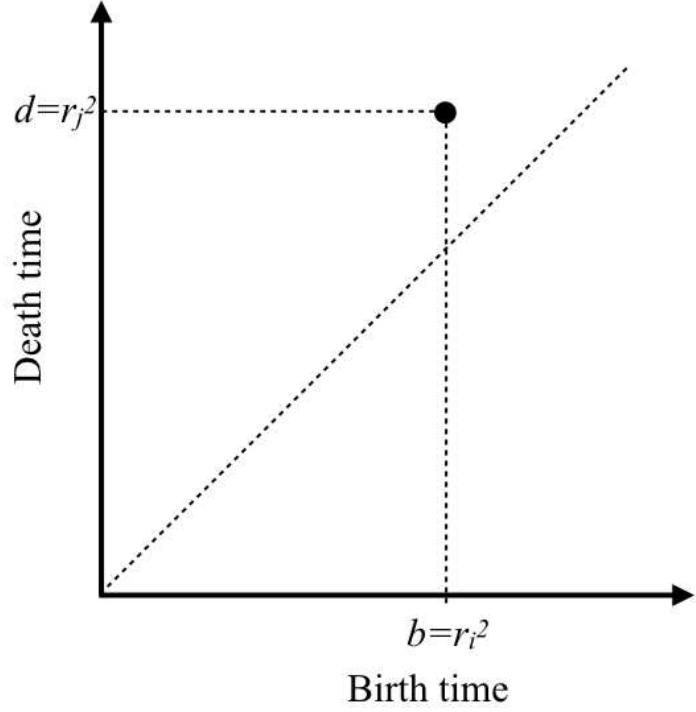}
 \caption{A schematic figure of persistent diagram (PD).
 The horizontal axis is the birth time $b=r_i^2$, and the vertical axis is the death time $d=r^2_j$
 }
 \label{PD}
\end{figure}

As $r^2$ varies, the process of changing of contents of complex is
called ``filtration''.  (According to the terminology of
persistent homology, one calls $r^2$ ``time'', but
the ``time'' is not real time.
It is simply measuring an effective size of the overlap of balls.)
The filtration process can be regarded as a coarse graining of the structure of data points.
We can understand the multiscale structure of data space by seeing the filtration process.
In particular, the existence of holes in the filtration process is important in the context of homology.
As already been illustrated in simple examples, intuitively,
the persistent homology counts holes created via the filtration.
A hole was born in Fig.~\ref{Filtration} (b) and died in Fig.~\ref{Filtration} (c).
Then, as shown in Fig.~\ref{PD}, we should plot a persistent diagram (PD)
whose horizontal and vertical axis are the birth and death times; the
birth time $b=r_i^2$ denotes the creation time of the holes in the filtration
and the death time $d=r_j^2$ does the disappearing time of corresponding
holes.

If the ratio of the birth time to the death one is approximately 1, we
should consider the hole having the birth and death time set $(b,d)$ as
a noise because its life time $d-b$ is almost zero.  Thus, we may pick
up the nontrivial correlation between data points from the persistent
diagram,
namely, by seeing birth-death-time sets far from the diagonal line in Fig.~\ref{PD}.
The hole which has a long life time is ``persistent'', hence, meaningful.

\subsection{Technical details of persistent homology}
\label{SubSec:PH2}

Let us express finite data points as $P=\{x_i \in \mathbb{R}^N ~|~
i=1,\cdots,m\}$ in the $N$-dimensional space where $m$ is the number of
data points.  For details of the computation of the persistent homology,
see Ref.~\cite{obayashi2017persistence} as an example.

To discuss the topological structure of the data points, one interesting
way is the calculation of the persistent homology~\cite{frosini1990distance,robins1999towards,zomorodian2001computing} and
then we need the geometric model~\cite{edelsbrunner1994three} which
characterized multiscale topological properties in $P$.  One of the
models is the $r$-ball model
\begin{align}
P_r &= \bigcup_{i=1}^m B_r (x_i),
\end{align}
where $B_r(x_i) = \{ y \in \mathbb{R}^N : ||y-x_i|| \le r \}$ and $r$
means the radius of each ball and its center is $x_i$.  The condition of
the crossing of two balls is expressed as
\begin{align}
 B_r(x_{i}) \cap B_r(y_{j}) \neq \emptyset,
 \label{two-balls}
\end{align}
where $\emptyset$ denotes an empty set.  Similarly, the condition of the
crossing of ($k+1$)-balls is expressed as
\begin{align}
 \bigcap_{j = 0}^{k}B_r(x_{i_j}) \neq \emptyset.
 \label{k-balls}
\end{align}
The {\v{C}}ech complex ${\cal }(P,r)$ is a nerve of the $r$-ball
collection $\Phi =\{B_r(x_i) | i=1,\cdots, m\}$ and is defined by
\begin{eqnarray}
{\cal C}(P,r) =\left\{ \{x_{i_{0}},\cdots,x_{i_k}\} \left| \bigcap_{j=0}^{k}B_r(x_{i_j})\neq \emptyset,
~~~k=0,\cdots,m-1
\right. \right\},
\label{chech}
\end{eqnarray}
where the set $\{x_{i_{0}},\cdots,x_{i_k}\}$ represents the
$k$-dimensional simplex which has the points $x_{i_{0}},\cdots,x_{i_k}$
as vertexes.  For example, $\{x_{i_0},\cdots,x_{i_k}\}$ for $k=1,2,3$ is
a line segment, a triangle and a tetrahedron, respectively.  In
Eq.~(\ref{chech}), for convenience, we have defined
\begin{eqnarray}
\left\{ \{x_{i_{0}}\} \left| \bigcap_{j=0}^{0}B_r(x_{i_j})\neq \emptyset
\right. \right\}
\equiv \left\{ \{x_{1}\}, \cdots,\{ x_{m}\} \right\},
\label{chech0}
\end{eqnarray}
which corresponds to the collection of vertex points themselves.  Note
that the ($k+1$)-balls crossing corresponds to the $k$-dimensional
simplex.  Hence, $k$ can be greater than the dimension $N$ of data
space, when the number $m$ of data points is greater than $N+1$.

\begin{figure}[h]
 \centering
 \includegraphics[width=0.7\textwidth]{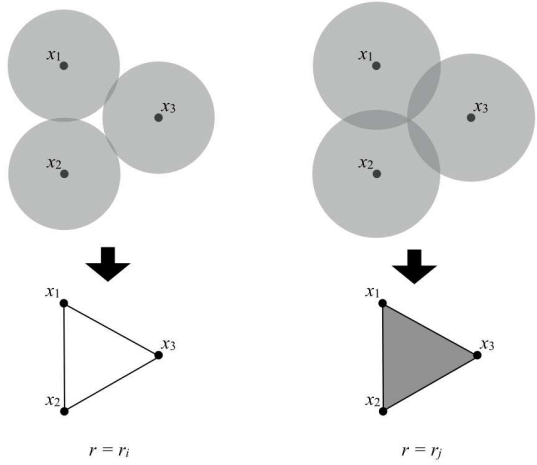}
 \caption{A schematic figure of the $r$-ball model (top) and the corresponding {\v{C}}ech complex (bottom) for 2-dimensional data space.}
 \label{Cech}
\end{figure}

Simple examples of \v{C}ech complex are shown in Fig.~\ref{Cech}. In the
left panel, the {\v{C}}ech complex is given by
\begin{eqnarray}
{\cal C}(P,r_i) =\left\{ \{x_1\},\{x_2\},\{x_3\},\{x_1,x_2\},\{x_2,x_3\},\{x_3,x_1\}\right\}
\label{fig1le}
\end{eqnarray}
In this case, there is no 2-dimensional simplex, namely, a (shaded)
triangle and there is a hole surrounded by 0 or 1-dimensional simplexes,
namely, points and line segments.  The right panel shows the case with
larger $r$.  The {\v{C}}ech complex is given by
\begin{eqnarray}
{\cal C}(P,r_j) =\left\{ \{x_1\},\{x_2\},\{x_3\},\{x_1,x_2\},\{x_2,x_3\},\{x_3,x_1\},\{x_1,x_2,x_3\}\right\}
\label{fig1re}
\end{eqnarray}
In this case, the (shaded) triangle appears and the hole vanished.
(Note that, in this case, the simplex with dimension higher than $N=2$
never appears for any $r$, since $m=3$ is not greater than $N+1=3$. )
As is clearly seen in the samples above, in general, the following
relation holds true.
 \begin{eqnarray}
{\cal C}(P,r)\subset {\cal C}(P,r^\prime ),~~~~~r<r^\prime.
\label{subset}
\end{eqnarray}
By using this relation, the filtration (increasing sequence) for the
{\v{C}}ech complex is now defined as
\begin{eqnarray}
 {\cal C}(P,r_0) \subset \cdots \subset {\cal C}(P,r_t) \subset \cdots \subset {\cal C}(P,r_{t_\mathrm{max}}),
\label{filtration}
\end{eqnarray}
where $r_0<\cdots <r_t<\cdots <t_{\rm max}$ and the filtration time $t$
is defined by $t=r^2$.  In principle, the time $t$ is a continuous real
number.  However, since the data space itself is discretized in actual
calculations, $r$ and $t$ are also treated as discretized quantities.
According to the nerve theorem, the nerve of $\Phi =\{B_r(x_i) |
i=1,\cdots, m\}$, namely, ${\cal C}(P,r)$ is homotopy equivalent with
$P_r$.  Hence, the filtration of ${\cal C}(P,r)$ has the information of
the multiscale topological properties, namely, the persistent homology,
in $P$.  The filtration process can be regarded as a coarse graining.
By the effects of the coarse graining, a topological structure (hole)
appears at some time $t$ (at some scale $r$) and disappears at larger
time (at larger scale).  For example, in Fig.~\ref{Cech}, we see that
the hole was born at the birth time $t_i\equiv {r_i}^2$ and died at the
death time $t_j\equiv {r_j}^2$.

To know the filtration, we need to check the crossing of multi balls.
Checking the crossing of two balls following the above condition
(\ref{two-balls}) is easy in the numerical computation.  However, the
numerical cost to check more than the crossing of three balls is large.
In the case of simple $r$-ball model, the maximum of the dimension of
the simplex which appears in the filtration process becomes larger as
$m$ (the number of data points) increases, and the computation time to
analyze the persistent homology becomes very large.  Thus, we employ the
alpha complex in this study as the more convenient version of the
$r$-ball model as usual in the persistent homology analysis.

To determine the alpha complex, we first assign the region $V(x_i)$ to the data
points as
\begin{align}
 V(x_i) = \{ x \in \mathbb{R}^N |~ ||x -x_i || \le ||x-x_j||,
 ~1 \le j \le m,~ j \neq i \},
\end{align}
and then we have $\mathbb{R}^N = \bigcup_{i=1}^m V(x_i)$.  The region,
$V(x_i)$, is so called the Voronoi region.  In the actual calculation,
we consider the Voronoi region consisted of $P$.  Next, we consider the
intersection defined by
\begin{align}
W_r(x_i) = B_r(x_i) \cap V(x_i).
\end{align}
It means that we restrict balls, $B_r(x_i)$, in the corresponding
Voronoi regions.  The alpha complex $\alpha(P,r)$ is defined by the
nerve of the collection $\Psi = \{W_r(x_i)|~i = 1, \cdots, j\}$, namely,
\begin{eqnarray}
\alpha(P,r) =\left\{ \{x_{i_{0}},\cdots,x_{i_k}\} \left| \bigcap_{j=0}^{k}W_r(x_{i_j})\neq \emptyset ,~~~k=0,\cdots m-1 \right. \right\}.
\label{alpha}
\end{eqnarray}

A necessary and sufficient condition for the existence of the
$k$-simplex in the alpha complex $\alpha(P,r)$ is
\begin{align}
 \bigcap_{j = 0}^{k}W_r(x_{i_j}) \neq \emptyset.
\label{condition}
\end{align}
If the condition mentioned above is satisfied for $r$, it holds true for
$r^\prime$ satisfying $r^\prime >r$.  Hence we get
\begin{align}
 \alpha(P,r) \subset \alpha(P,r'),~~ r < r'.
\end{align}
As is in the case of \v{C}ech complex, the filtration for the alpha
complex is defined as
\begin{align}
 \alpha(P,r_0) \subset \cdots \subset \alpha(P,r_t) \subset \cdots \subset \alpha(P,r_{t_\mathrm{max}}),
\end{align}
where $r_0 < \cdots < r_t < \cdots < r_{t_\mathrm{max}}$.

\begin{figure}[h]
\centering
 \includegraphics[width=0.9\textwidth]{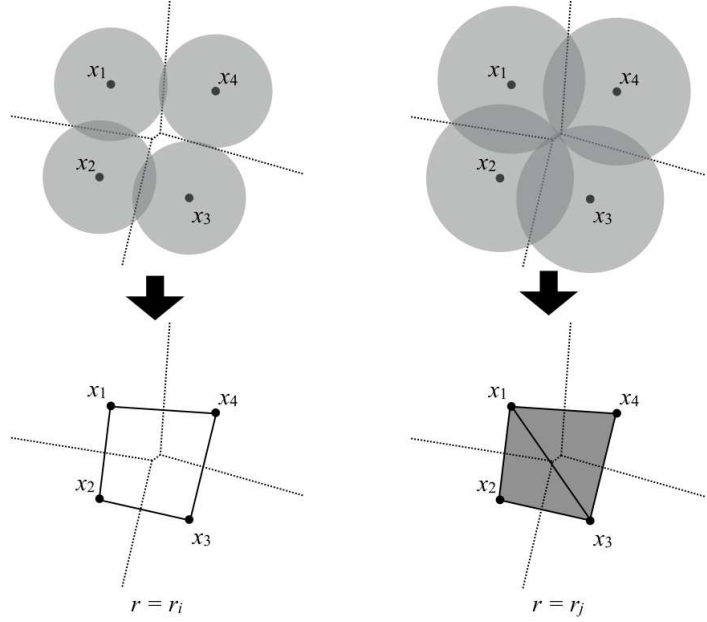}
 \caption{A schematic figure of the filtration of the alpha complex for two dimensional data space.
 The left (right) panel shows the alpha complex $\alpha(P,r_i)$ ($\alpha(P,r_j)$) ($r_i < r_j$).
 The dashed lines denote the borders of the Voronoi region.
 In the left panel, the hole surrounded by $W_r(x_i)$ was born at the birth time $b = r_i^2$.
 In the right panel, the hole disappears at the death time $d = r_j^2$.
 }
 \label{fig:demo-alpha}
\end{figure}

A schematic figure of the filtration of alpha complexes are shown in
Fig.~\ref{fig:demo-alpha}.  In the left panel, the alpha complex is
given by
\begin{eqnarray}
\alpha(P,r_i) =\left\{ \{x_1\},\{x_2\},\{x_3\},\{x_4\}, \{x_1,x_2\},\{x_2,x_3\},\{x_3,x_4\},
\{x_4,x_1\} \right\}
\label{alphal}
\end{eqnarray}
In this case, there are only 0 or 1-dimensional simplexes (points and
line segments) , and a hole surrounded by them is formed.  The right
panel shows the case with larger $r$.  In this case, the alpha complex
is given by
\begin{eqnarray}
\alpha(P,r_j)& =&\left\{ \{x_1\},\{x_2\},\{x_3\},\{x_4\}, \{x_1,x_2\},\{x_2,x_3\},\{x_3,x_4\},
\right.
\nonumber\\
&&\left.
\{x_4,x_1\}, \{x_1,x_3\},\{x_1,x_2,x_3\},\{x_1,x_3,x_4\}
\right\}
\label{alphar}
\end{eqnarray}
The 2-dimensional simplexes (shaded triangles ) appears and the hole
vanishes.  The hole appears at the birth time $b=r_i^2$ and disappears
at the death time $d={r_j}^2$.

It should be remarked that 3-dimensional simplex (tetrahedron)
\begin{eqnarray}
\{x_1,x_2,x_3,x_4\},
\label{tetra}
\end{eqnarray}
which have a larger dimension than the dimension $N=2$ of data space
itself, does not appears in the right panel of Fig.~\ref{fig:demo-alpha}
(also in Eq.~(\ref{alphar})).  The relation $W_r(x_2)\cap
W_r(x_4)=\emptyset $ holds true and the 3-dimensional simplex never
appears even if $r$ becomes larger than $r_j$, since $V(x_2)$ is not
neighboring to $V(x_4)$.  In general, the alpha complex defined for
$N$-dimensional data space does not have simplexes the dimension of
which is greater than $N$, when there is no special point $x$ which
satisfies
\begin{eqnarray}
||x-x_{i_1}||=\cdots =||x-x_{i_j}||=||x-x_{i_{N+2}}||.
\label{general-position}
\end{eqnarray}
This situation makes the computation of persistent homology~\cite{frosini1990distance,robins1999towards,zomorodian2001computing} in
alpha complex much easier than that in {\v{C}}ech complex.  Actually, by
using this filtration of alpha complex, we can calculate the persistent
homology .  Actual construction of the alpha complex in the numerical
simulation, we employ DIPHA (Distributed Persistent Homology Algorithm)
library~\cite{dipha} via the homcloud-base software~\cite{homCloud}.

\section{Effective Polyakov-line model}
\label{Sec:EPL}
One of the interesting QCD effective models is the effective
Polyakov-line (EPL) model~\cite{Greensite:2014isa,Greensite:2014cxa}.  The grand canonical
partition function~\cite{Hirakida:2017bye} is
\begin{align}
{\cal Z } &= \int {\cal D} U~e^{-(S_\mathrm{G} + S_\mathrm{Q})}, \\
 S_\mathrm{Q} &= \sum_{{\mathbf x}} \mathcal{L}_\mathrm{Q} (\mathbf{x}),
 \nonumber\\
 S_\mathrm{G} &= - \kappa \sum_{{\mathbf x}} \sum_{k=1}^3
 \Bigl(
 \mathrm{Tr} [U_\mathbf{x}]
 \mathrm{Tr}[U^\dag_{\mathbf{x}+{\hat k}}]
 + \mathrm{c.c.}
 \Bigr),
\end{align}
where $U_\mathbf{x}$ means the Polyakov-line holonomy defined by
\begin{align}
 U_\mathbf{x} =
 \mathrm{diag}(e^{i\phi_1},e^{i\phi_2},e^{-i(\phi_1+\phi_2)}),
\end{align}
with the real parameters $\phi_1$ and $\phi_2$, and the coupling
constant $\kappa$ in $S_\mathrm{G}$ relates with the inverse temperature
$\beta=1/T$.  Note that $S_{\mathrm{G}}$ and $S_{\mathrm{Q}}$ correspond
to the gluon and quark actions, respectively, in QCD.
Since we can consider a variety of the EPL model and thus we choose
logarithmic $\mathcal{L}_\mathrm{Q}$ in this study as
\begin{align}
 \mathcal{L}_\mathrm{Q}
 = &- \ln \Biggl[ \mathrm{det}
 \Bigl\{ 1 + e^{-\beta M} U_\mathrm{x} \Bigr\}
 ^{2 N_\mathrm{f}}
 \mathrm{det}
 \Bigl\{ 1 + e^{-\beta M}U_\mathrm{x}^\dag \Bigr\}
 ^{2 N_\mathrm{f}}
 \Biggr],~
\end{align}
where $M$ is the quark mass and $N_\mathrm{f}=3$ is the number of
flavors.
In this model, $M$ and $T$ should depend on the coupling constant $\kappa$.
However,
it is difficult to take the line of constant physics unlike the lattice QCD
simulation, and we need to solve the effective theory to obtain the relation according
to Refs.~\cite{Greensite:2014isa,Greensite:2014cxa}.
In this study, for qualitative study, we treat $\beta M = M/T$ as a parameter
independent of $\kappa$,
and consider the high or the low $T$ regime by varying $\kappa$.

Usually, the confinement-deconfinement nature in this model is
characterized by the configuration average value $\langle |L| \rangle$
of the spatial averaged Polyakov-line operator
\begin{align}
 L = \dfrac{1}{V}\sum_\mathrm{x} \dfrac{1}{3}\mathrm{Tr}[U_\mathrm{x}],
\end{align}
where $V$ is the three-dimensional volume; small $\langle |L| \rangle$,
$\langle |L| \rangle \sim 0$, indicates the confined phase and large
$\langle |L| \rangle$, $\langle |L| \rangle \sim 1$, does the deconfined
phase.  $L$ is not invariant under $\mathbb{Z}_3$ transformation, but is
not an exact order parameter since $S_\mathrm{Q}$ breaks the
$\mathbb{Z}_3$ symmetry explicitly.

\begin{figure}[h]
 \centering
 \includegraphics[width=0.7\textwidth]{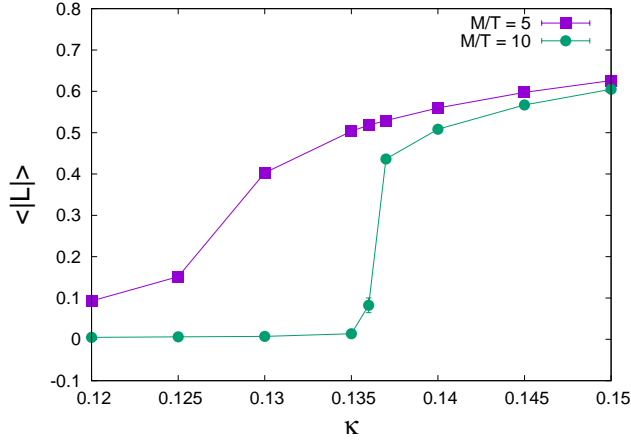}
 \caption{The $\kappa$-dependence of $\langle |L| \rangle$ with $M/T=5,~10$.}
 \label{fig:ploop_kappa-dep}
\end{figure}

\begin{figure}[h]
 \centering
 \includegraphics[width=0.48\textwidth]{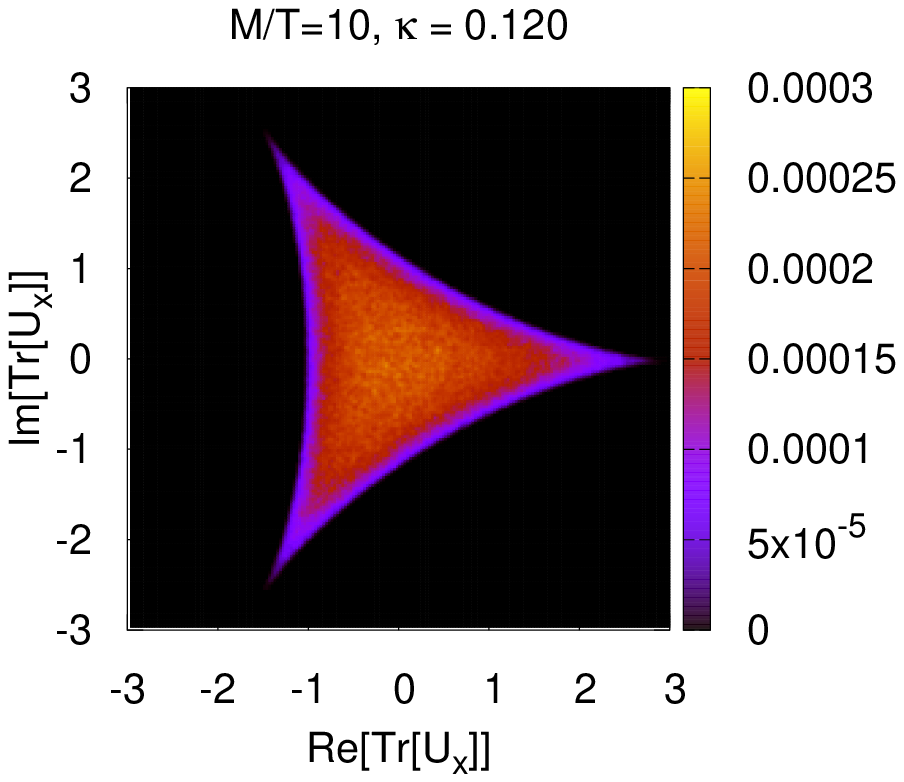}
 \includegraphics[width=0.48\textwidth]{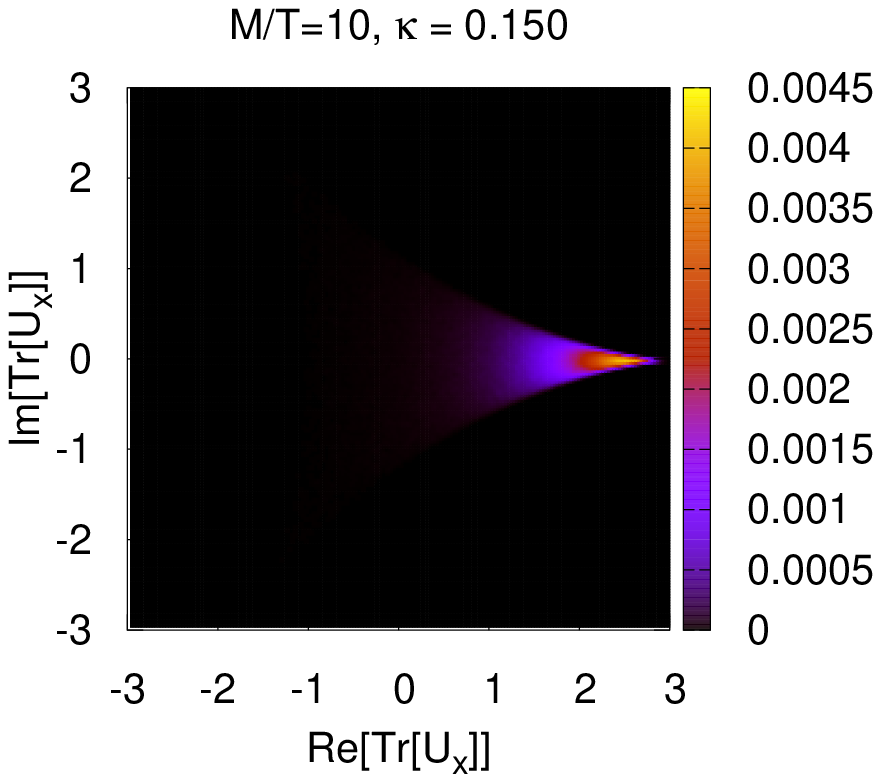}
 \caption{The probability density of $\mathrm{Tr}[U_\mathrm{x}]$ in the complex plane.
 The left and right panels show the results for $\kappa=0.120$ and $\kappa=0.150$ with $M/T=10$,
 respectively.}
 \label{fig:pb-ploop}
\end{figure}

To perform the path integral of the EPL model, we use the Monte Carlo
method; we generate configurations to replace the integral by the
statistical sum.  For the numerical simulation, we consider $V = 24^3$,
$\kappa \in [0.120,0.150]$, and two cases with $M/T = 5$ and $10$.  For
reader's convenience, we here show the $\kappa$-dependence of $\langle
|L| \rangle$ in Fig.~\ref{fig:ploop_kappa-dep} where configuration
average is taken by $50$ configurations as an example.  We can clearly
see that this model exhibits the same tendency as QCD and there should
be the phase transition between $\kappa =0.135$ and $0.140$.
Figure~\ref{fig:pb-ploop} shows the probability density of $\mathrm{Tr}[U_\mathrm{x}]$
in the complex plane.
At $\kappa=0.120$ (confined phase), the probability density is widely spread in the plane.
On the other hand, at $\kappa=0.150$ (deconfined phase), the density is mainly distributed
around the real axis and $\mathrm{Re}[\mathrm{Tr}[U_\mathrm{x}]] \gtrsim 0$.

\section{Numerical results}
\label{Sec:Results} In
Refs.~\cite{Gattringer:2010ms,Borsanyi:2010cw,Endrodi:2014yaa,Schafer:2015wja},
it is shown that the number of the percolating cluster (which has a same
scale as the spatial volume) is 3 (1) in the confined (deconfined)
phase, and the expectation value of the weight for the largest cluster
normalized by the spatial volume is small in the confined phase and it
rapidly grow up when the temperature becomes larger than the critical
temperature.  They indicate that the large structures exist at both
phases.  However, these results also indicate the possibility that the
large hole structures may not appear in the deconfined phase since the
largest cluster dominates too strongly.  Therefore, by using the
persistent homology, it is expected that we can extracted the large
holes in the confined phase, while many tiny holes appear in the
deconfined phase.

\begin{figure}[h]
 \centering
 \includegraphics[width=0.7\textwidth]{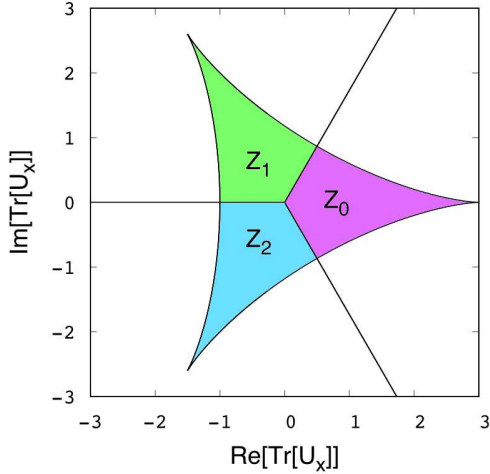}
 \caption{A schematic figure of Polyakov-line in the complex plane.
 To compute the persistent homology, we divide the plane into three $\mathbb{Z}_3$ domains,
 ${\cal Z}_0$, ${\cal Z}_1$, and ${\cal Z}_2$.
 }
 \label{fig:Ploop_comic}
\end{figure}

To analyze the spatial distributions of simulation data via the
persistent homology analysis, we consider the following isolation
procedure:
\begin{enumerate}
 \item Map the bare data for each site to the complex
       Polyakov-line plane, configuration by configuration.
 \item Divide the complex Polyakov-line plane into three domains as shown
       in Fig.~\ref{fig:Ploop_comic}.
       Note that the ${\cal Z}_1$ and ${\cal Z}_2$ domains are the
	   $\mathbb{Z}_3$-images of the ${\cal Z}_0$ domain.
 \item Prepare three lists,
       ${\cal Z}_0,{\cal Z}_1$, and ${\cal Z}_2$,
       as data storage places.
       The storage places, ${\cal Z}_0,{\cal Z}_1$, and ${\cal Z}_2$,
       become the $N^3$ dimensional list since we try to maintain
       the spatial information of the bare data.
 \item Storage the mapped data to each list; if the data are located in
       the ${\cal Z}_0$ domain in Fig.~\ref{fig:Ploop_comic}, the
       corresponding site is ON (full) in the ${\cal Z}_0$ group and the
       corresponding sites are OFF (empty) in the ${\cal Z}_{1,2}$
       groups.
\end{enumerate}

\begin{figure}[h]
 \centering
 \includegraphics[width = 0.4\textwidth]{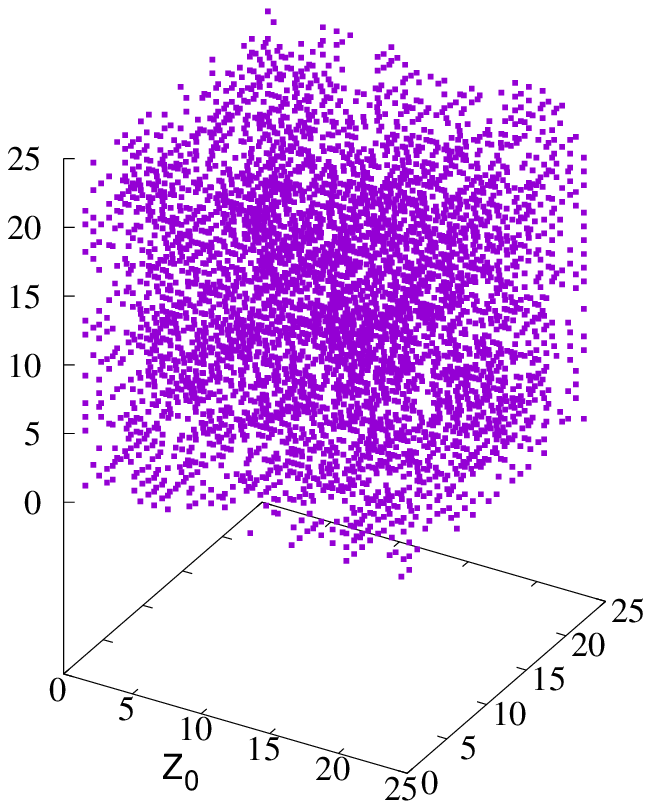}
 \includegraphics[width = 0.4\textwidth]{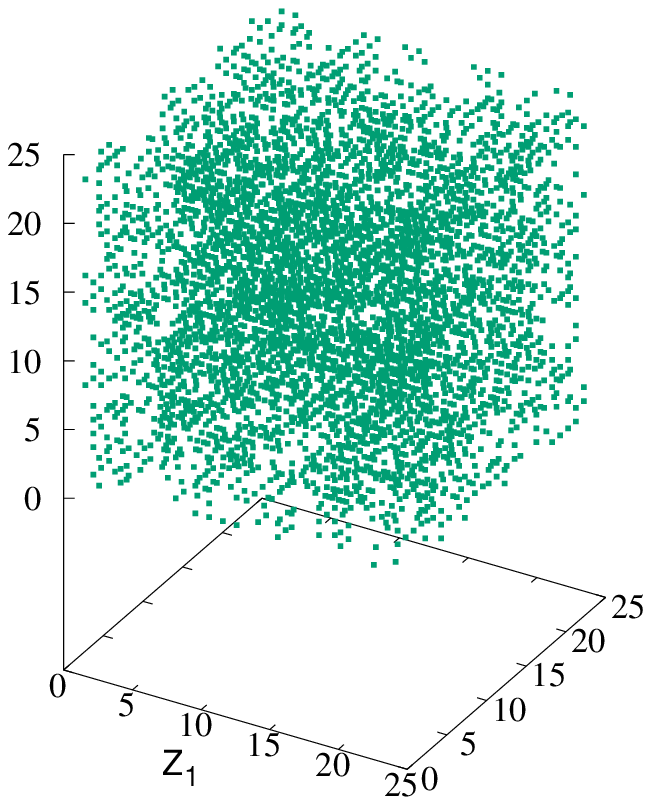}
 \includegraphics[width = 0.4\textwidth]{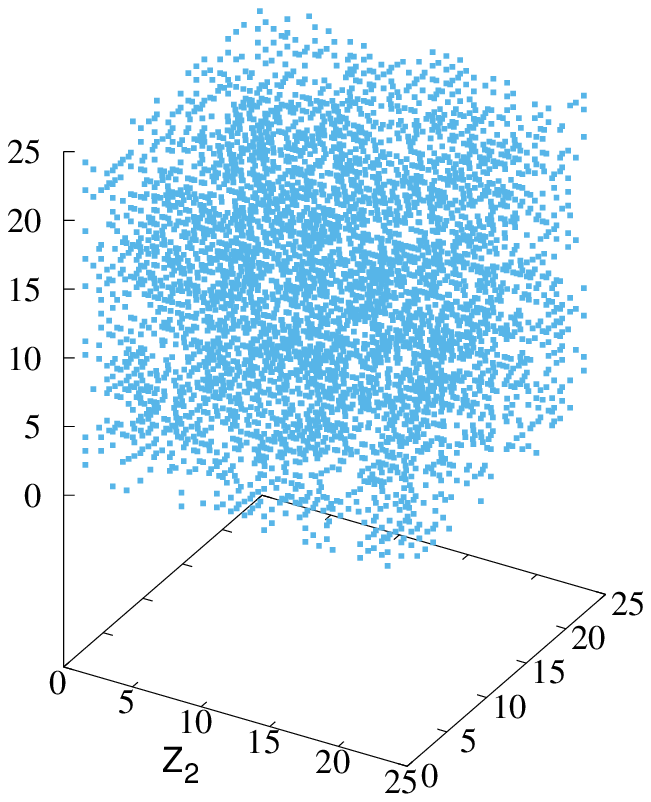}
 \caption{The spatial distributions of the data after the isolation for $\kappa = 0.120$ and $M/T=10$.
 The top left, top right and bottom data show the mapped data of ${\cal Z}_0$, ${\cal Z}_1$ and ${\cal Z}_2$.}
 \label{fig:map_data_k0120}
\end{figure}

\begin{figure}[h]
 \centering
 \includegraphics[width = 0.4\textwidth]{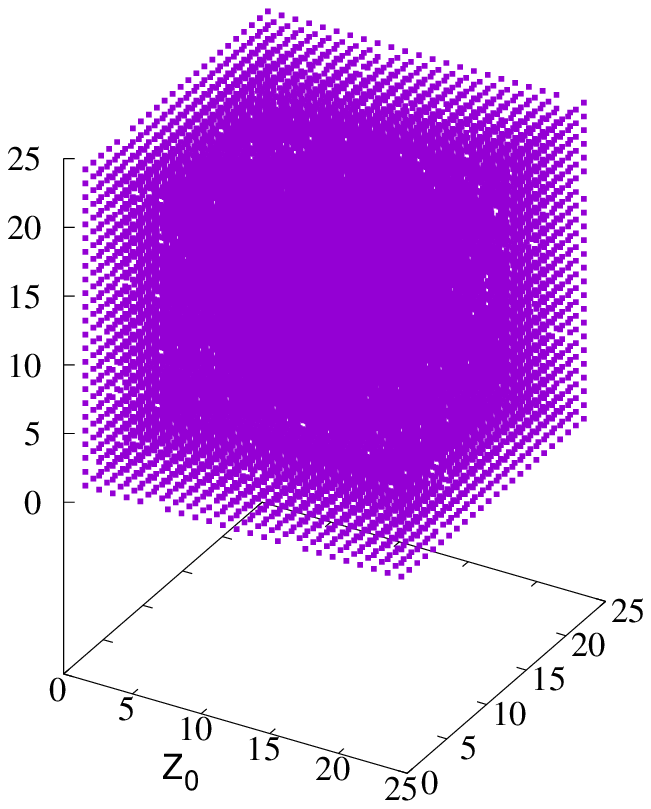}
 \includegraphics[width = 0.4\textwidth]{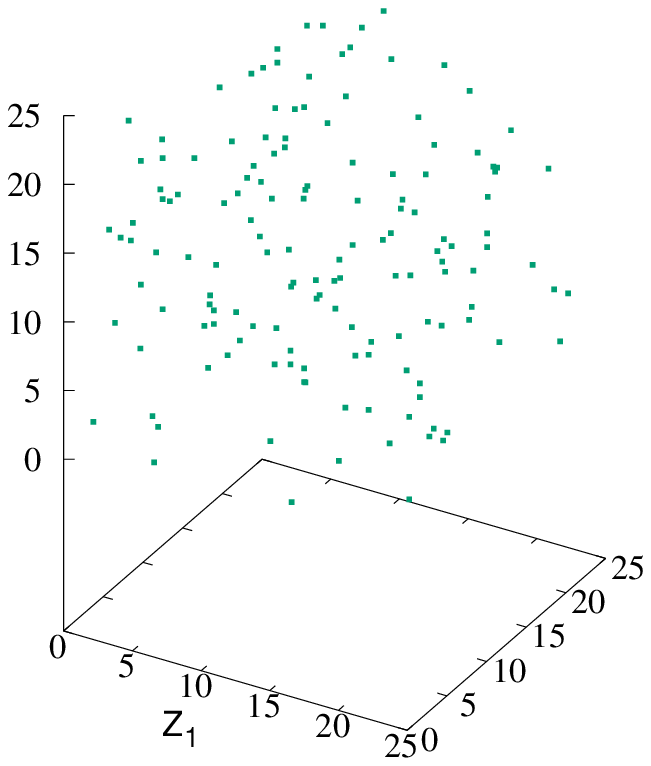}
 \includegraphics[width = 0.4\textwidth]{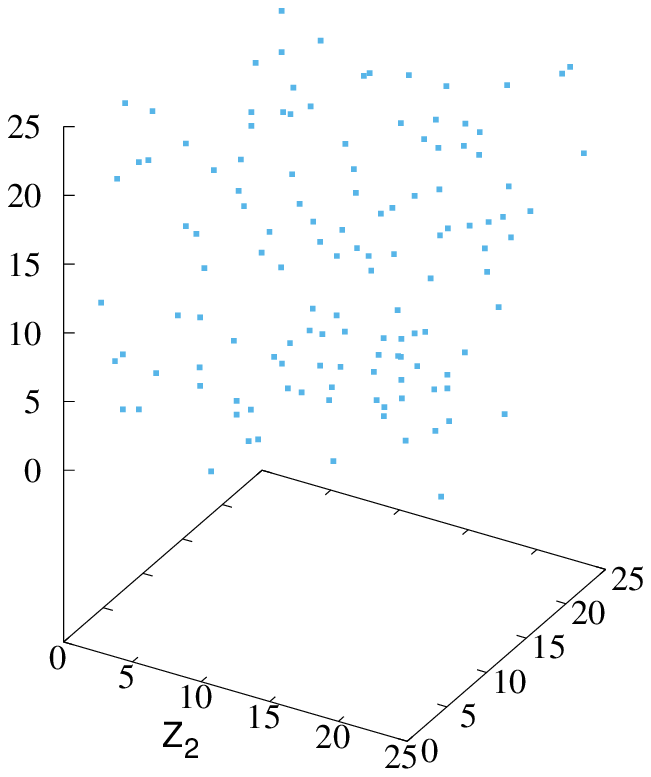}
 \caption{The spatial distributions of the data after the isolation for $\kappa = 0.150$ and $M/T=10$.
 The top left, top right and bottom data show the mapped data of ${\cal Z}_0$, ${\cal Z}_1$ and ${\cal Z}_2$.}
 \label{fig:map_data_k0150}
\end{figure}

First we consider the spatial distribution of the data after the above
isolation procedure in the case with $M/T=10$.  The typical examples are
shown in Fig.~\ref{fig:map_data_k0120} and Fig.~\ref{fig:map_data_k0150}
for $\kappa=0.120$ and $\kappa=0.150$, respectively.  At $\kappa=0.120$
(confined phase), the data are uniformly distributed, but the
distribution is weighted toward the ${\cal Z}_0$ domain and few data are
located in the ${\cal Z}_1$ and ${\cal Z}_2$ domains at $\kappa=0.150$
(deconfined phase).

\begin{figure}[h]
 \centering
 \includegraphics[width=0.8\textwidth]{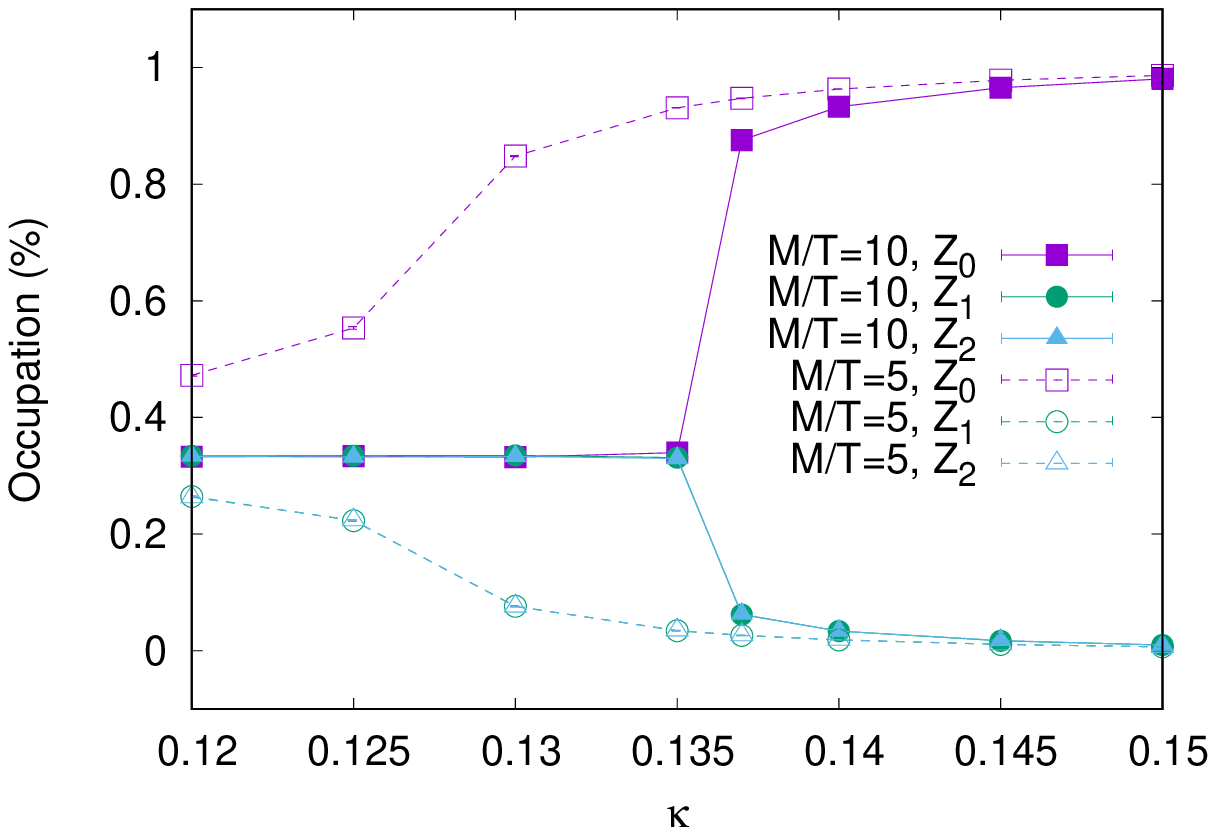}
 \caption{The occupation rate of each domain in the lattice space.
 The solid (dashed) lines shows the results in the case with $M/T=10$ ($M/T = 5$).
 For each case, the square, circle and triangle symbols denote the occupation rate of ${\cal Z}_0$, ${\cal Z}_1$, and ${\cal Z}_2$ domain, respectively.
 Here, we define the occupation rate as the ratio of
 the number of data points in each domain to the lattice volume $V$.}
 \label{fig:occ-Z}
\end{figure}

Figure~\ref{fig:occ-Z} shows the occupation rates of each domain in the
lattice space.
Here, we define the occupation rate as the ratio of
the number of
data points in each domain to the lattice volume $V$.
For the calculation of the rates, we take a
configuration average with 50 configurations.  In the case with $M/T =
10$, the occupation rates of all domains are almost the same (about
$33\%$) below $\kappa = 0.135$, while the ${\cal Z}_0$ domain occupies
almost all lattice sites ($90 \sim 99\%$) in the deconfined phase.
These behaviors are almost the same with the results shown in
Ref.~\cite{Gattringer:2010ms}.  On the other hand, in the case with $M/T
= 5$, the rate of ${\cal Z}_0$ domain is larger by about $11\%$ than
that of other domains at $\kappa = 0.120$, and the rate also increases
as $\kappa$ becomes large.  This difference between the two cases
indicates that the explicit $\mathbb{Z}_3$ symmetry breaking is not
negligible when $M/T = 5$.

From these situations, it is expected that the persistent diagram for
all domains are almost the same in the confined phase, while the result
in the ${\cal Z}_0$ domain of deconfined phase differs from that of
other domains due to the difference in the occupation rate.  If the
center clusters with nontrivial topological structures exist in the
lattice space of each domain, the persistent diagram may be different
from that of data points distributed randomly.  The comparison of the
results of the EPL model with that of the random distributed
configuration is discussed in the end of this section.

\begin{figure}[h]
 \centering
 \includegraphics[width = 0.45\textwidth]{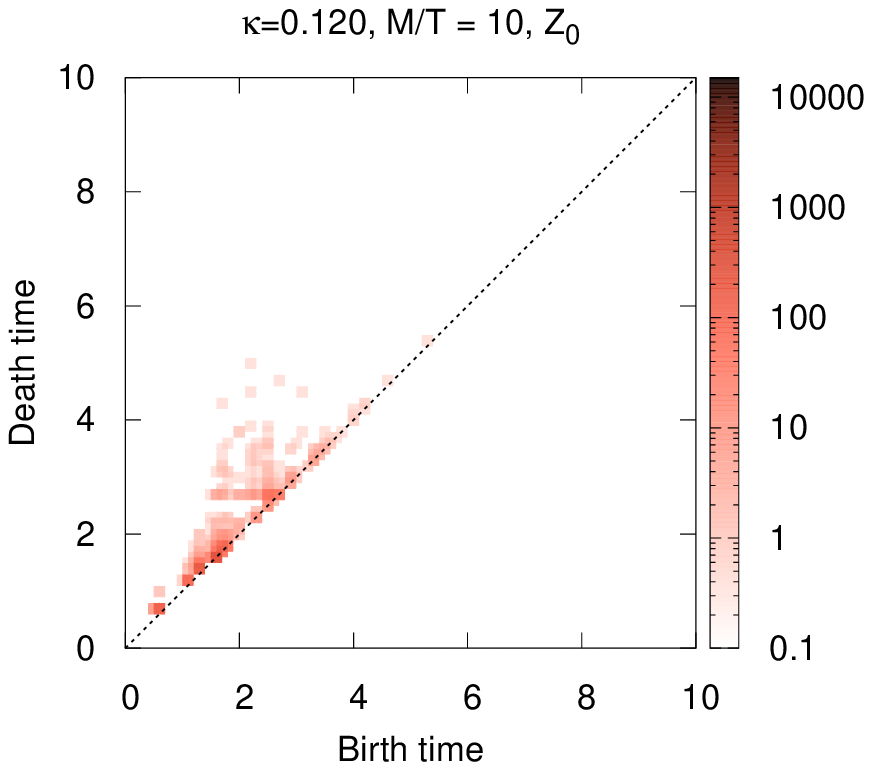}
 \includegraphics[width = 0.45\textwidth]{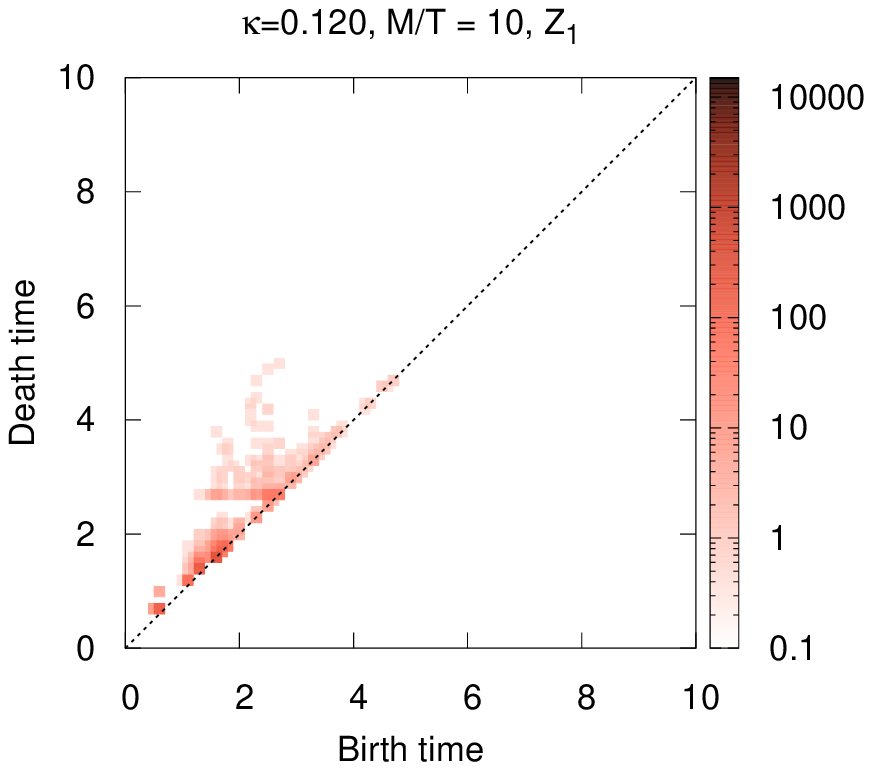}
 \includegraphics[width = 0.45\textwidth]{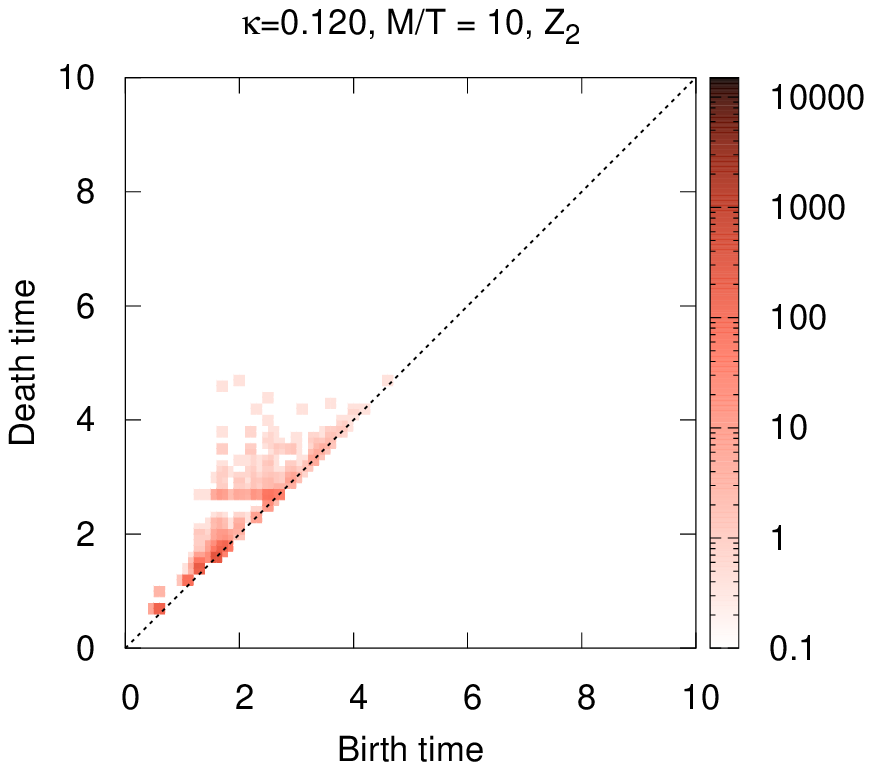}
 \includegraphics[width=0.45\textwidth]{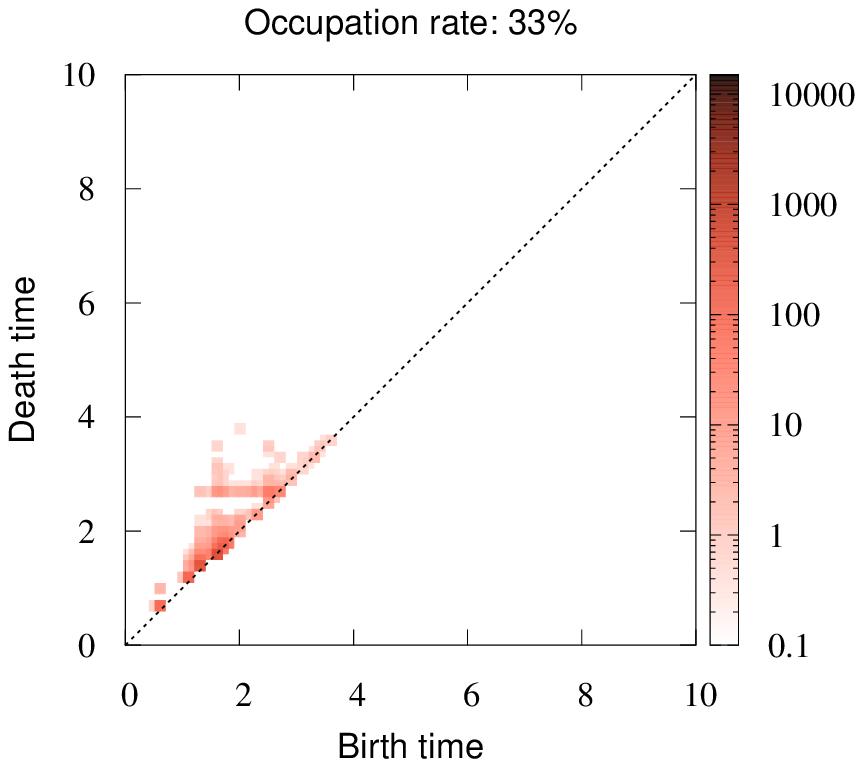}
 \caption{The persistent diagram for $\kappa = 0.120$.
 The top left, top right, and bottom panels show the result of ${\cal Z}_0$, ${\cal Z}_1$, and ${\cal Z}_2$ list, respectively.
 The diagram of the random configuration
 with the occupation rate $33\%$ is also shown.
 The color bar expresses the creation number of each hole.}
 \label{fig:pd_k0120}
\end{figure}
\begin{figure}[h]
 \centering
 \includegraphics[width = 0.45\textwidth]{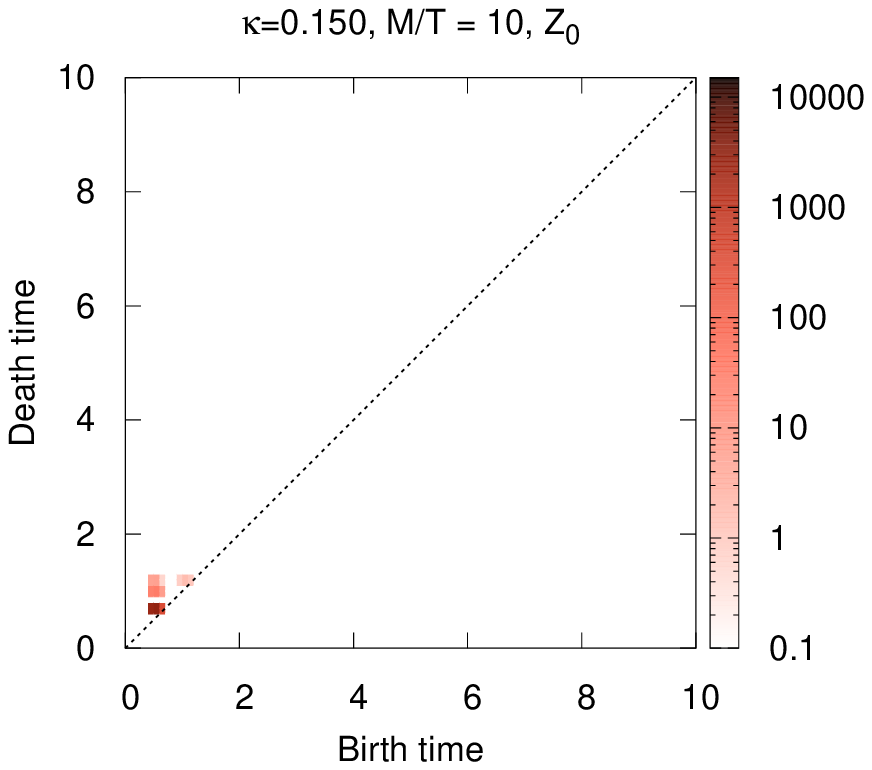}
 \includegraphics[width = 0.45\textwidth]{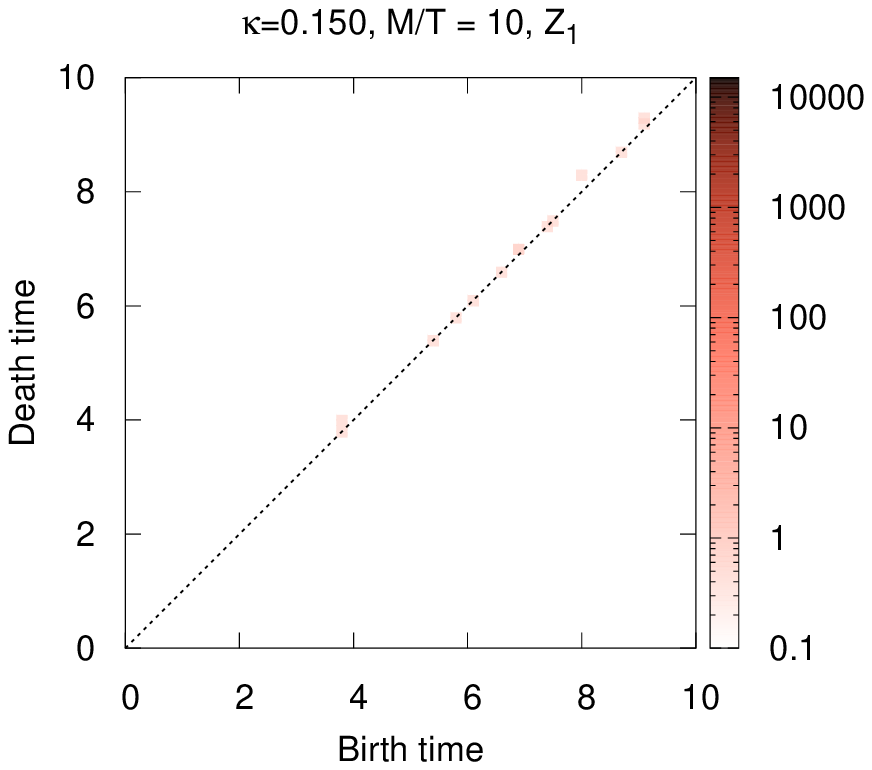}
 \includegraphics[width = 0.45\textwidth]{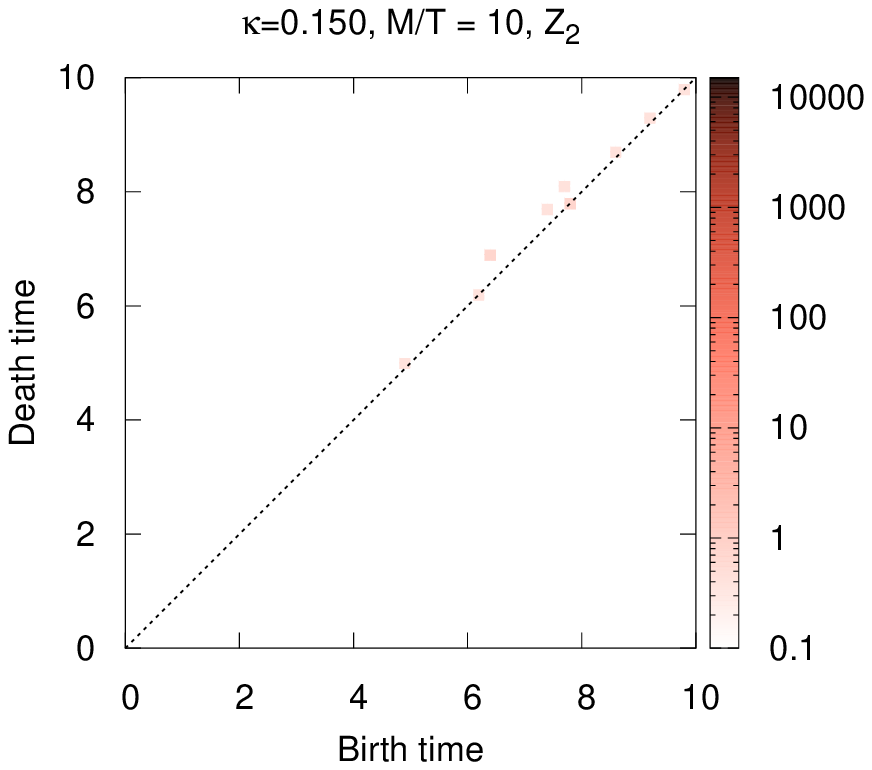}
 \includegraphics[width=0.45\textwidth]{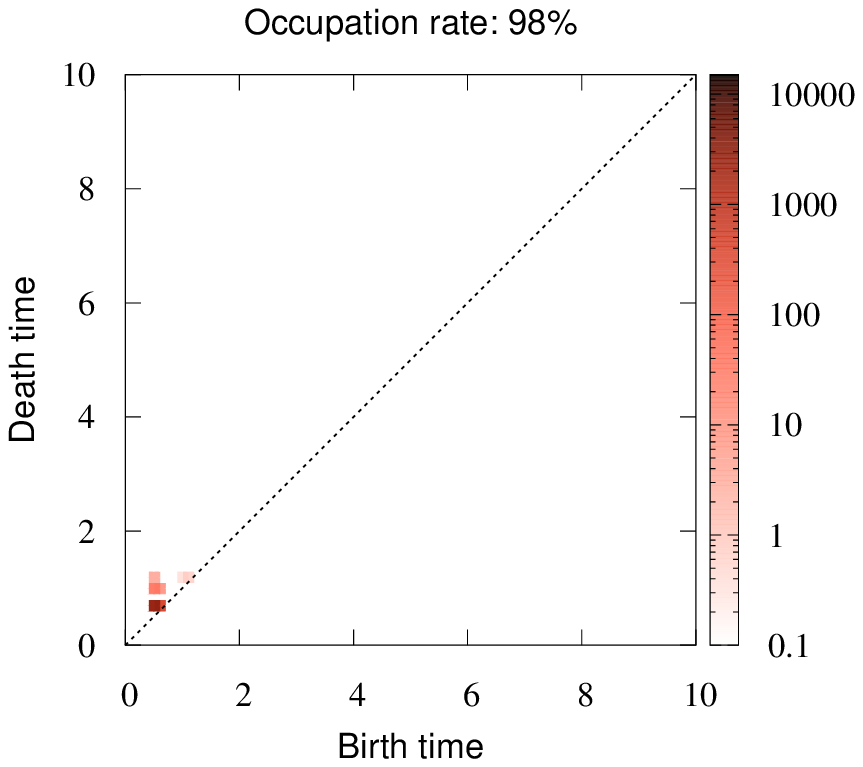}
 \caption{The persistent diagram for $\kappa = 0.150$.
 The top left, top right, and bottom panels show the result of ${\cal Z}_0$, ${\cal Z}_1$, and ${\cal Z}_2$ list, respectively.
 The diagram of the random configuration
 with the occupation rate $98\%$ is also shown.
 The color bar expresses the creation number of each hole.}
 \label{fig:pd_k0150}
\end{figure}

Now we analyze the data of the EPL model by the method of the persistent
homology.  In this, paper, since we are interested in 3-dimensional
structure of the data, we search a 3-dimensional hole which is
surrounded by 2-dimensional surfaces.  In other words, we search the
surfaces of the polyhedron the vertexes of which are data points.  The
typical examples of the persistent diagram are shown in
Fig.~\ref{fig:pd_k0120} and Fig.~\ref{fig:pd_k0150} with $\kappa=0.120$
and $0.150$, respectively.  The horizontal and the vertical axes are the
birth time $b_i$ and the death time $d_i$ of the $i$-th hole,
respectively.

\begin{figure}[h]
 \centering
 \includegraphics[width=0.8\textwidth]{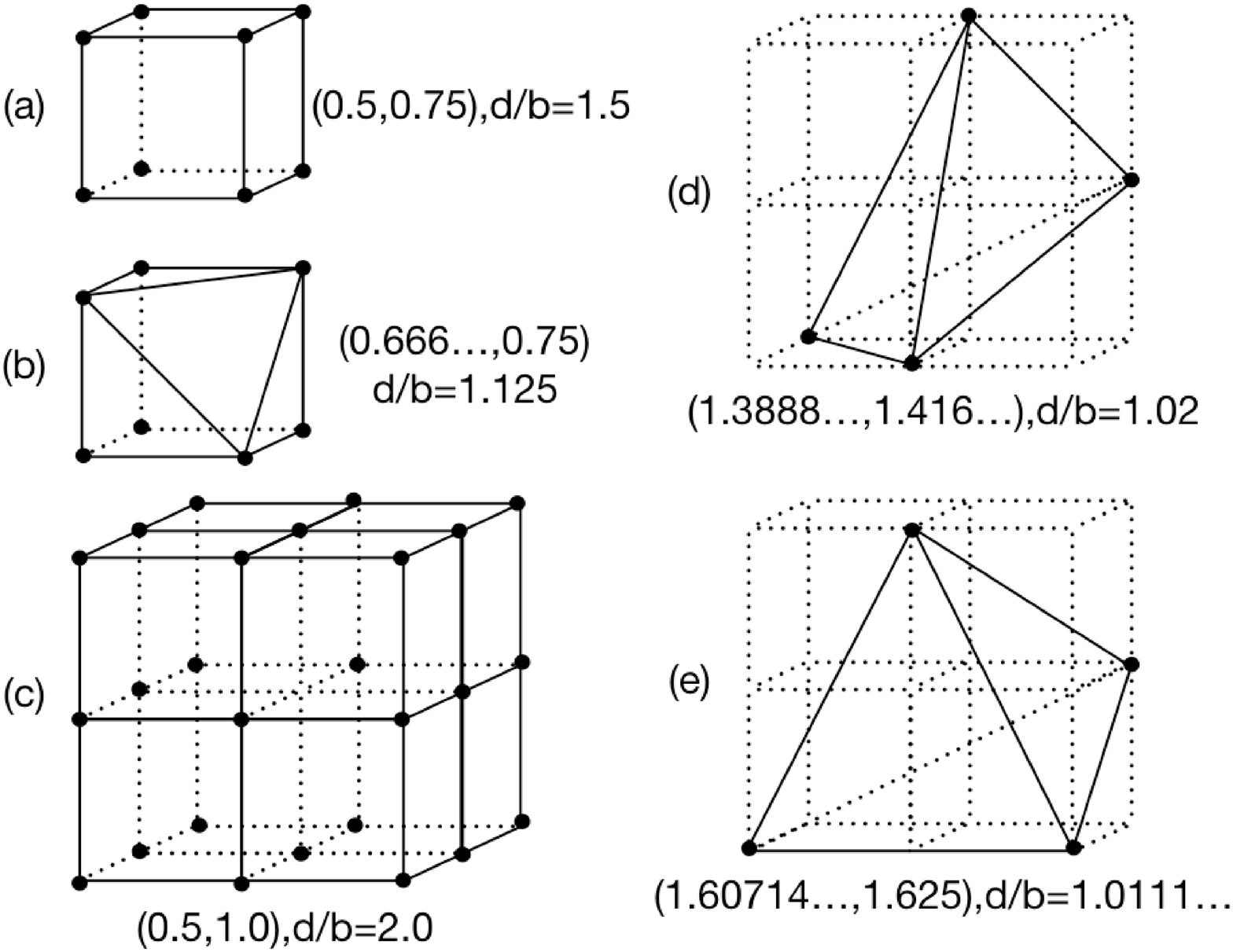}
 \caption{Typical shapes appearing in the persistent homology analysis
 on the rectangular lattice.
 In the set $(b,d)$, $b$ and $d$ mean the birth and death time,
 respectively.
 }
 \label{fig:ph_comic}
\end{figure}

Figure~\ref{fig:ph_comic} summarizes typical shapes of data points in
the persistent homology analysis with the set of these birth and death
times.
These shapes appear as dominant structures in the domains.  In this
study, we call the most generated, the second most generated, and the
third most generated shapes as ``dominant structures'' or
``typical shapes'' for the domain.  The shape (a) is the smallest cube
on the present rectangular lattice and (b) is the chipped cube of (a).
The shape (c) is the $2\times 2\times 2$ cube but it does not contain
the data point at its center.  Both (d) and (e) are cubes which can
appear in the rectangular parallelepiped.  In these shapes, (a), (b),
(d) and (e) can be expected to appear as dominant holes of mapped data
and thus those creation numbers become large; for example, since the
shape (c) can be considered as the composite of eight (b), eight (b) are
always created if one (c) is created.


Let us consider the most simple structure (a) for the example of
calculating the birth-death-time set $(b,d)$.  This is the smallest cube
in the lattice space.  When the radius of the sphere at the vertex of
the cube is $r = \sqrt{1^2 + 1^2}/2 = 1/\sqrt{2}$, the half of the
diagonal of the face, a hole surrounded by six faces is created.  After
increasing the radius, the hole disappears when the radius is $r' =
\sqrt{1^2 + 1^2 + 1^2}/2 = \sqrt{3}/2$, the half of the diagonal of the
cube.
Thus, we obtain
the time of birth and death for
the structure (a) as $(b,d) = (r^2, r'^2) = (0.5,0.75)$.

The time of birth and death for the other structures in Fig.~\ref{fig:ph_comic} are
also obtained by the same calculation: The birth time is the square of the half of the
longest diagonal of the face of the polyhedron corresponding to the
hole, while the death time is the square of the half of the longest
diagonal of the polyhedron.
Based on the idea mentioned above, we can determine the typical shapes in Fig.~\ref{fig:ph_comic}
inversely from their time of birth and death.


At $\kappa=0.120$, the number of holes and their variety of shapes are
almost same for each list, since $M/T$ is large and $\mathbb{Z}_3$
symmetry is approximately preserved.  In this phase, the
birth-death-time sets, $(1.3888\cdots,1.41666\cdots)$,
$(0.666\cdots,0.75)$ and $(1.60714\cdots,1.625)$, dominate the
persistent diagram and these time sets are corresponding to the shape
(d), (b) and (e) in Fig.~\ref{fig:ph_comic}, respectively.  The shapes
(d) and (e) are seemed to be constructed, because the occupation rate of
each domain is about 33\% and the distance between data points is
moderately far.

In comparison, at $\kappa=0.150$, mapped data are weighted toward the
${\cal Z}_0$ domain and thus many holes which have the fast birth and
death times are created.  However, mapped data are very sparse in the
${\cal Z}_1$ and ${\cal Z}_2$ domains, and thus we need long time to
create holes and then the death times are not much larger than the birth
times, hence, the life times $d_i-b_i$ of holes become very short.  This
fact can be seen from Fig.~\ref{fig:pd_k0150}.  In the deconfined phase,
the birth-death-time sets, $(0.5,0.75)$, $(0.666\cdots,0.75)$ and
$(0.5,1.0)$, dominate the persistent diagram for the ${\cal Z}_0$ list
and these time sets are corresponding to the shape (a), (c) and (b) in
Fig.~\ref{fig:ph_comic}.  In the ${\cal Z}_1$ and ${\cal Z}_2$ lists, a
few holes are created and then the mapped data are very sparse, and thus
we do not show typical shape shown in Fig.~\ref{fig:ph_comic}.  In this
phase, the ${\cal Z}_0$ domain contains almost all lattice sites, so
that the shape (a), the simplest and smallest structure in the lattice
space, dominates in the persistent diagram. But this is a trivial
structure.  For the shapes (b) and (c), it is considered that they
appear as minute structures if some data points belong to other domains.
In particular, the one shape (c) with eight shapes (b) are constructed
if the $2 \times 2 \times 2$ cube doesn't contain the data point at its
center.  However, we may not be able to investigate these tiny
structures in detail.  To check them, it is needed to analyze in the
larger lattice volumes, or restrict the Polyakov-line phase in smaller
regions as in the percolation analysis in
Refs.~\cite{Gattringer:2010ms,Borsanyi:2010cw,Endrodi:2014yaa,Schafer:2015wja}.
Dominant structures in each phase are summarized in Table~\ref{tab:dominant}.
\begin{table}[h]
 \centering
 \caption{Dominant structures for each phase.}
 \begin{tabular}[t]{|c||c|l|}
  \hline
  Phase & Group & Dominant structure \\ \hline
  \multirow{3}{*}{Confinement} & \multirow{3}{*}{${\cal Z}_0$,${\cal Z}_1$,${\cal Z}_2$}
      & (d), $(1.3888\cdots,1.41666\cdots)$ \\
  & & (b), $(0.666\cdots,0.75)$ \\
  & & (e), $(1.60714\cdots,1.625)$ \\ \hline
  \multirow{4}{*}{Deconfinement} & & (a), $(0.5,0.75)$ \\
  & ${\cal Z}_0$& (b), $(0.666\cdots,0.75)$ \\
  & & (c), $(0.5,1.0)$ \\ \cline{2-3}
  & ${\cal Z}_1$, ${\cal Z}_2$ & No dominant structure \\ \hline
 \end{tabular}
 \label{tab:dominant}
\end{table}

\begin{figure}[h]
 \centering
 \includegraphics[width=0.8\textwidth]{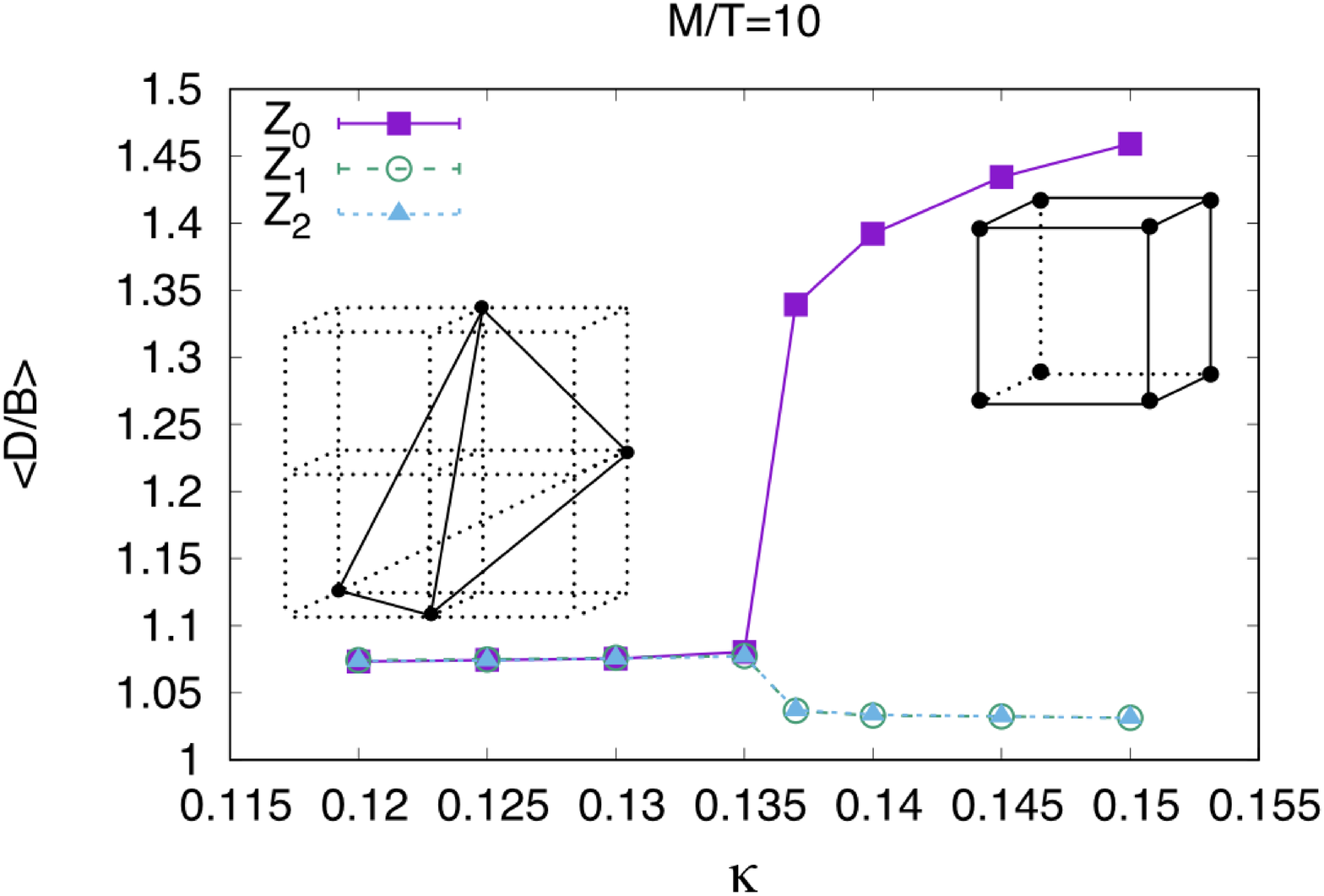}
 \caption{The $\kappa$ dependence of $\langle D/B \rangle$ for $M/T = 10$.
 Cube symbol shows the results of ${\cal Z}_0$, circle symbol shows the results of ${\cal Z}_1$,
 and triangle symbol shows the results of ${\cal Z}_2$. The shapes on
 the figure represent the typical ones which dominate the values in the
 confined and the deconfined phases.}
 \label{fig:ph_death-birth_MT-10}
\end{figure}

Next, we analyze the $\kappa$-dependence of the persistent homology with
$M/T=10$.  However, the individual persistent diagram is not convenient
in this case and thus we introduce the averaged ratio of the birth and
death times as
\begin{align}
 \label{eq:d-to-b}
 D/B &\equiv \dfrac{1}{N_\text{hole}}\sum_{i}^{N_\text{hole}}
 \dfrac{d_i}{b_i},
\end{align}
where $N_\text{hole}$ is the number of holes in each list.  In this
study, we calculate $D/B$ except for $d_i/b_i < 1.001$, since the hole
with the ratio $d_i/b_i$ close to 1 is considered as noise.  We here
take the statistical average $\langle D/B\rangle$ by using $50$
configurations and the result is shown in
Fig.~\ref{fig:ph_death-birth_MT-10}.  In the confined phase, $\langle
D/B\rangle$ always becomes $\sim 1.073$ for all lists.  In comparison,
$\langle D/B\rangle$ of the ${\cal Z}_0$ list is changed into $\sim
1.40$ and reaches $\sim 1.459$ at $\kappa = 0.150$, but $\langle
D/B\rangle$ of ${\cal Z}_1$ and ${\cal Z}_2$ decreases toward $\sim
1.03$ in the deconfined phase.  This characteristic behavior indicates
the shape change in each list by the confinement-deconfinement
transition.

It seems that the increase of $\langle D/B\rangle$ in deconfined phase
is mainly caused by the fact that the birth time is limited since the
largest percolating cluster dominates too
strongly~\cite{Gattringer:2010ms,Borsanyi:2010cw,Endrodi:2014yaa}, and
the trivial structure (a) dominates in large $\kappa$-limit. On the
other hand, the larger holes which has the larger birth time can exist
in confined phase, since the three percolating clusters dominate not so
strongly.

\begin{figure}[h]
 \centering
 \includegraphics[width=0.8\textwidth]{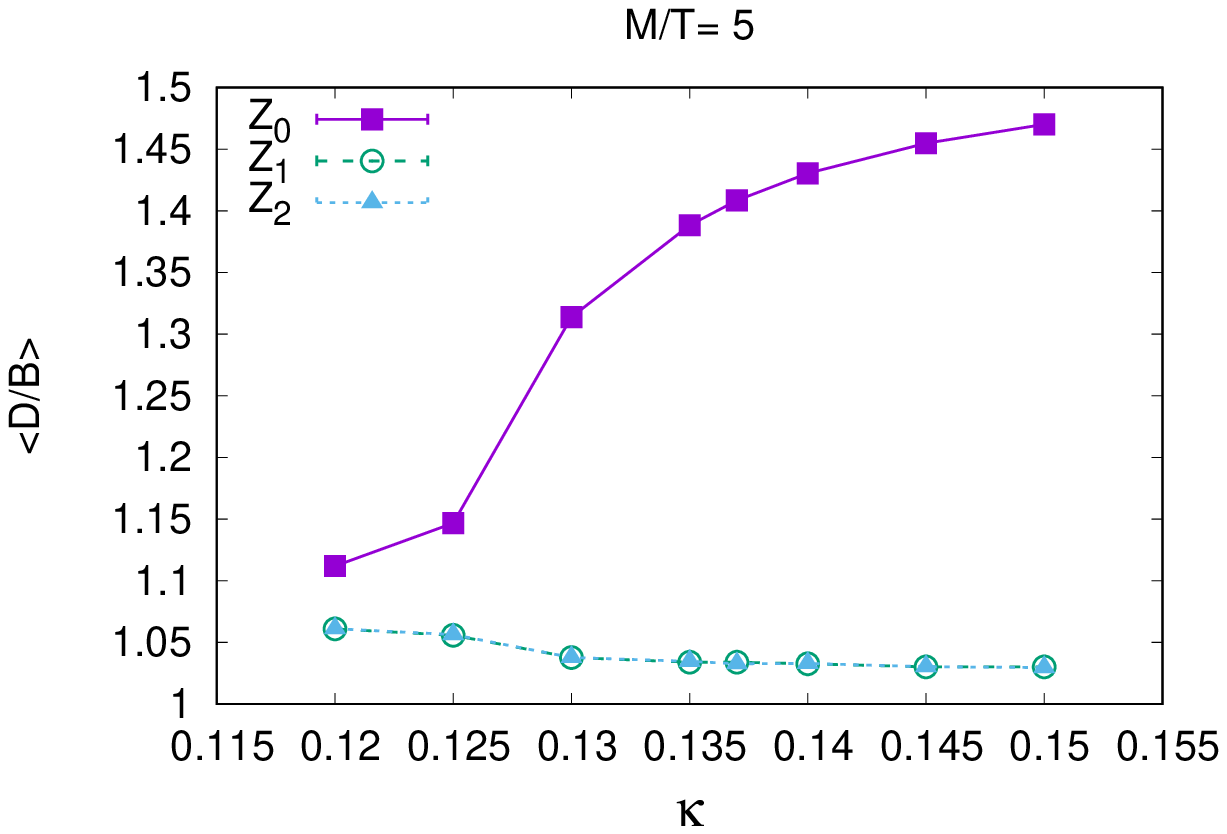}
 \caption{The $\kappa$ dependence of $\langle D/B \rangle$ for $M/T = 5$.
 Cube symbol shows the results of ${\cal Z}_0$, circle symbol shows the results of ${\cal Z}_1$,
 and triangle symbol shows the results of ${\cal Z}_2$.}
 \label{fig:ph_death-birth_MT-05}
\end{figure}

Figure~\ref{fig:ph_death-birth_MT-05} shows $\langle D/B\rangle$ with
$M/T=5$.  The explicit breaking of $\mathbb{Z}_3$ symmetry is enhanced
at small $\kappa$ due to the lighter quark mass, and the
$\kappa$-dependence of $\langle D/B\rangle$ becomes smooth, but
asymptotic values at large $\kappa$ are same as the results with $M/T=10$.

It should be noted that the statistic error becomes quite small for the
persistent homology analysis.
This results indicate that the distributions of data points
for all configurations in each list are very similar.


Let us compare the results of the EPL model with that of the
randomly distributed configuration.  Here, we generate the configuration
of data points in random so as to have an arbitrary occupation rate, and
analyze it with the persistent homology.
In the EPL model with $M/T = 10$, the occupation rates of all domain are
about 33\% at $\kappa = 0.120$, and that of the ${\cal Z}_0$ domain at
$\kappa = 0.150$ is about 98\%.
The bottom right panels in Figs.~\ref{fig:pd_k0120}
and \ref{fig:pd_k0150} show the typical persistent diagrams of the random configurations corresponding to
the occupation rate $98\%$.

Comparing the results of the EPL model (top left, top right, and
bottom left panels in Fig.~\ref{fig:pd_k0120})
with those of the random
configuration with the occupation rate $33\%$,
we find that the number of holes with the
death time $3.5 \lesssim d \lesssim 5$ in the random configuration is
much less than that in the EPL model at $\kappa = 0.120$ for $M/T = 10$.
This result is confirmed in all domains and all configurations.  In the
random configuration of the occupation rate 33\%, the dominant shapes
are (d), (e) and the shape with the birth-death-time set (1.125,1.25)
(which does not appear in Fig.~\ref{fig:ph_comic}), while the dominant
ones are (d), (e) and (b) in the EPL model.
As is already mentioned, the shape
(c) can be considered as the composite of eight (b).
If all of the shape (b) are used to construct the shape (c), the ratio of the number of (c)
to that of (b) becomes 1/8$=$ 0.125.
However, in our calculations, the ratio is 0,
and this result indicates that the shape (c) does not exist
but the shape (b) exists independently.
This property appears in $\kappa \le 0.130$ for $M/T = 10$.
The shape (b) has the chipped $1 \times 1 \times 1$ cube structure,
and this shape is likely to appear on the surface of the clusters of data points.
These results indicate existence of
the more complex multiscale topological structures
in the confined phase with $M/T=10$.

Meanwhile, comparing the results of the EPL model (top left, top right, and
bottom left panels in Fig.~\ref{fig:pd_k0150})
with those of the random configuration with the occupation rate $98\%$,
we find that both persistent diagrams are
almost the same.  This may be a trivial result because almost all
lattice sites are included in the ${\cal Z}_0$ domain at $\kappa =0.150$
for $M/T = 10$ and the large holes cannot be constructed.  Thus,
this result is consistent with the indication mentioned in the beginning
of this section.  The dominant structures of the random configuration
are also the same with that of the ${\cal Z}_0$ domain of the EPL model
at the deconfined phase.  For the $Z_0$ domain of deconfined phase, the
result of the EPL model is qualitatively consistent with that in the
random distribution model.

\begin{figure}[h]
 \centering
 \includegraphics[width=0.8\textwidth]{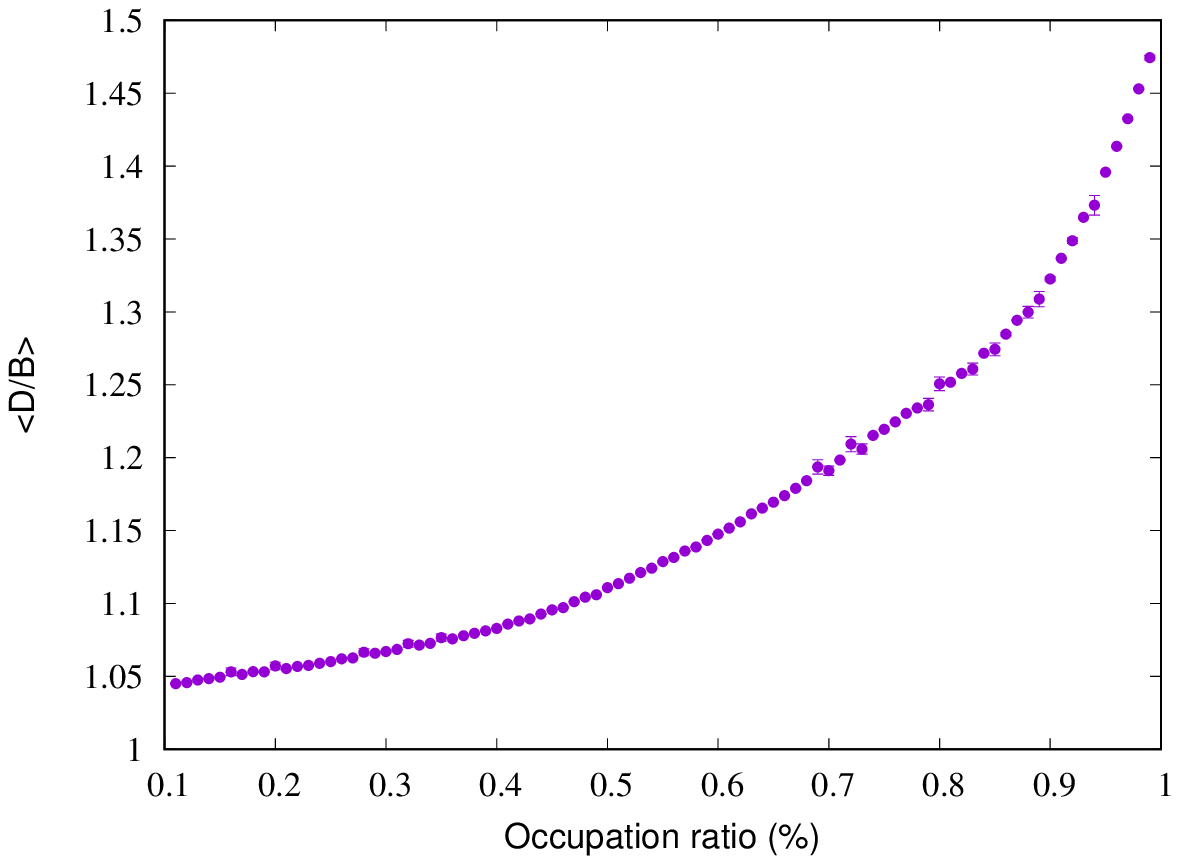}
 \caption{The occupation rate dependence of $\langle D/B \rangle$ for the random configuration.
 The statistical error is shown, but it is tiny.}
 \label{fig:random-db}
\end{figure}

To compare the EPL model with the random distribution model more
quantitatively, we compute $\langle D/B \rangle$ in the random model.
Figure~\ref{fig:random-db} shows the occupation rate dependence of
$\langle D/B \rangle$ for the random configuration.  To compute $\langle
D/B \rangle$ for each occupation rate, we take a statistical average
with 50 configurations.  As the occupation rate increasing, $\langle D/B
\rangle$ increases monotonously.

\begin{figure}[h]
 \centering
 \includegraphics[width=0.48\textwidth]{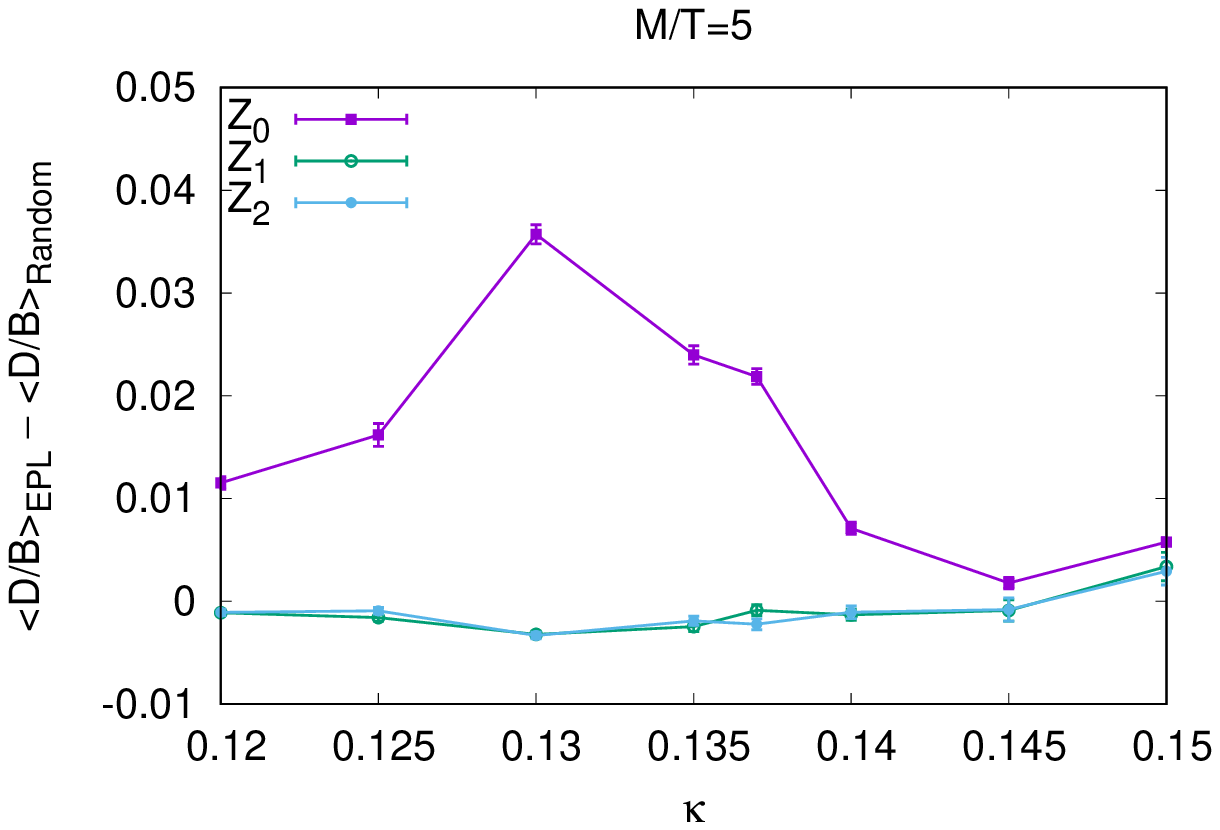}
 \includegraphics[width=0.48\textwidth]{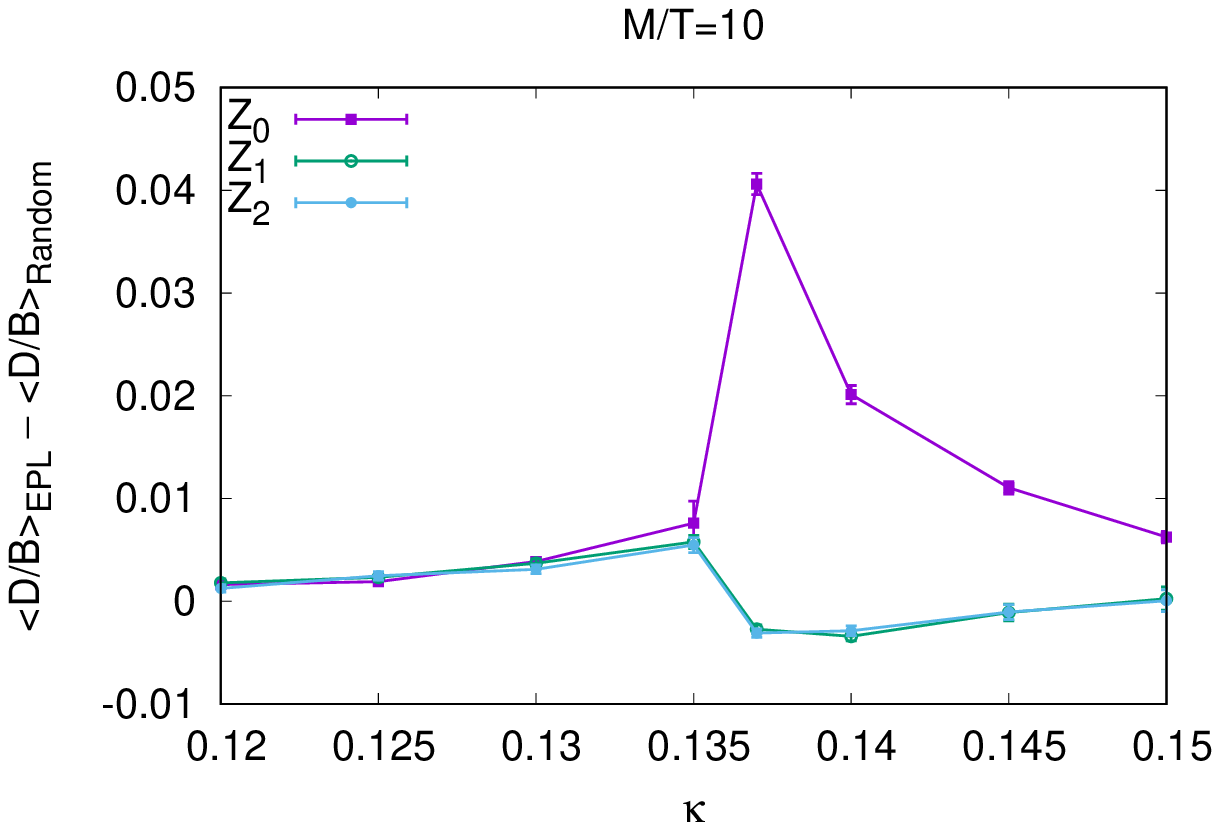}
 \caption{The differences of $\langle D/B \rangle$ between the EPL model
 and the random configuration.
 The left and right panels show the results with $M/T = 5$ and $M/T=10$,
 respectively.}
 \label{fig:db2}
\end{figure}

In Fig.~\ref{fig:db2}, we compare $\langle D/B \rangle$ in
the EPL model with that of the random configuration model with the same
occupation rate.
In the ${\cal Z}_0$ domain for both $M/T$ and in the ${\cal Z}_{1,2}$ domains in
$\kappa \le 0.135$ for $M/T = 10$ and $\kappa=0.150$ for $M/T=5$,
the value in the EPL model is larger
than that of the random configuration: for example, the value in the EPL
model at $\kappa = 0.120$ for $M/T = 10$ is larger than
$\langle D/B \rangle \sim 1.071$
for the occupation rate 33\% in the
random configuration model by $\sim 0.002$,
while the value in the ${\cal Z}_0$
at $\kappa =0.150$ is lager than
$\langle D/B \rangle \sim 1.452$
for the rate 98\% in the random configuration by $\sim 0.006$.
On the other hand, in the other domains and parameter sets, the values in the
EPL model are smaller than those of the random configuration.  These
difference may come from that the EPL model has the different
topological structures from the random one.  In particular, in the
${\cal Z}_0$ domain at $\kappa = 0.150$ for $M/T = 10$, $\langle D/B
\rangle$ is somewhat different from that of the random configuration,
although the persistent diagrams are almost the same in both the cases.
In addition, especially around $\kappa = 0.130$ for $M/T=5$
and $\kappa = 0.137$ for $M/T = 10$, the differences in the ${\cal Z}_0$ domain
become much large.
These results indicate that the ${\cal Z}_0$ domain with both $M/T$ has
much different structures topologically there.

\begin{figure}[h]
 \centering
 \includegraphics[width=0.48\textwidth]{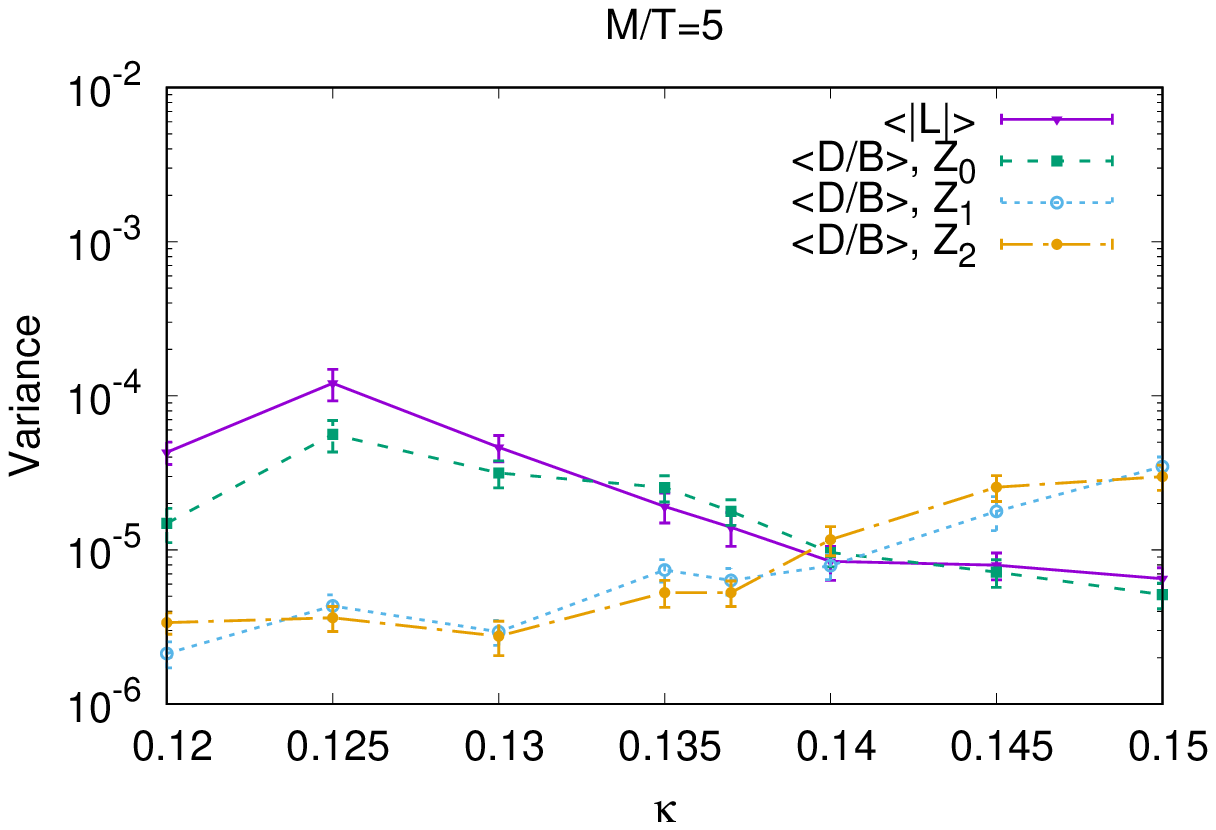}
 \includegraphics[width=0.48\textwidth]{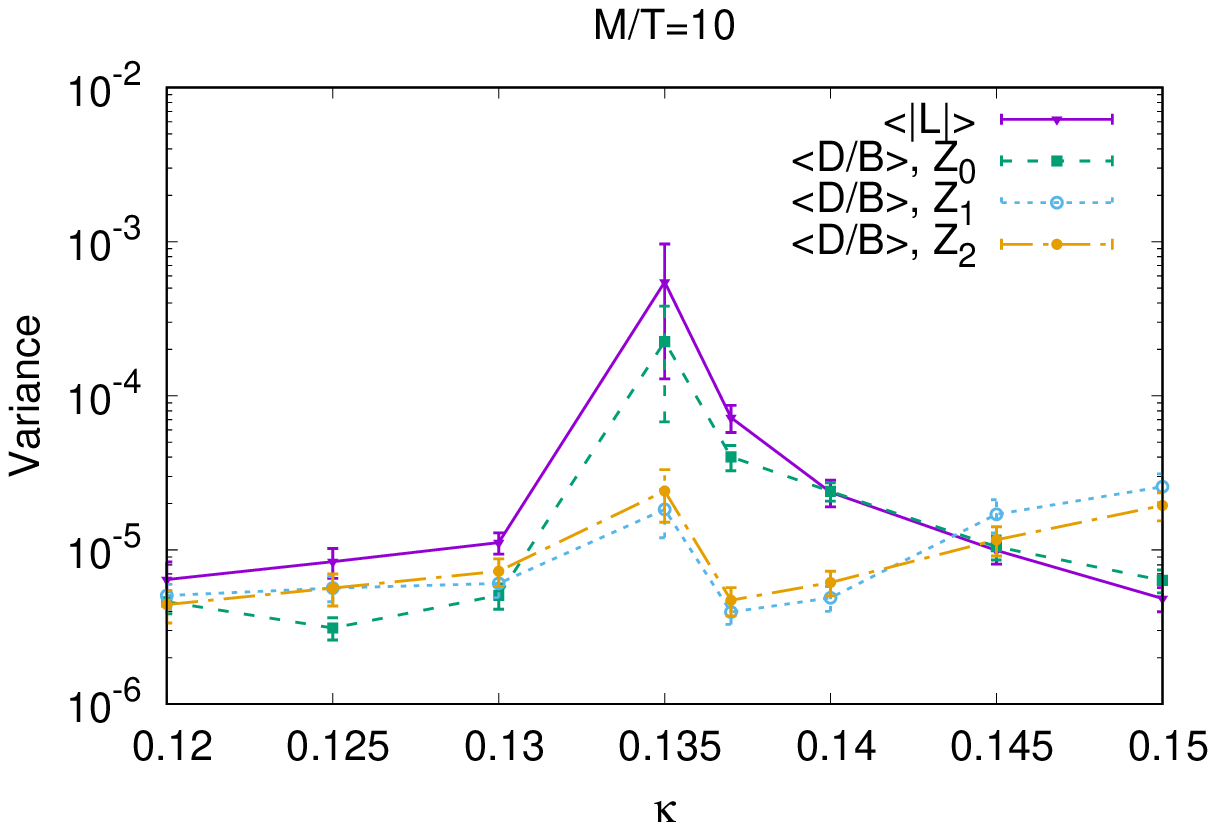}
 \caption{The comparison of the variance of $D/B$ and $|L|$.
 The left and right panels show the results with $M/T = 5$ and $M/T=10$, and
 the solid and dashed lines denote the results of $|L|$ and $D/B$ for each domain,
 respectively.}
 \label{fig:variance}
\end{figure}

In Fig.~\ref{fig:variance}, we compare the variance
$\langle O^2 \rangle - \langle O \rangle^2$ of $D/B$ and $|L|$.
For the results of $|L|$,
a peak is located at $\kappa \sim 0.135$
for $M/T=10$
and $\kappa \sim 0.125$
for $M/T=5$.
Meanwhile, the peaks in the case of $\langle D/B \rangle$ for ${\cal Z}_0$ domain
are located at almost the same $\kappa$ in the case of $|L|$,
while the variances for the other domains behave almost the same.
These results indicate that we can investigate properties of the confinement-deconfinement
transition with $D/B$.


From these results, by using $\langle D/B \rangle$, we have
characterized the phases.  This quantity is calculated after obtaining
the persistent diagram, and this diagram has the information of the
multiscale spatial structure of data in the entire lattice space.  Thus,
we can use $\langle D/B \rangle$ as the nonlocal order parameter
characterizing the confined and deconfined phases.  As mentioned
earlier, the statistical error of $\langle D/B \rangle$ is much smaller
than that of the expectation value of the Polyakov-line $\langle |L|
\rangle$.  This may be caused by the fact that the structures for each
configuration
have the similar spatial distribution.
Therefore, we can
characterize the phases with a fewer configurations than in the case of
calculating $\langle |L| \rangle$.


\section{Summary}
\label{Sec:Summary}
In this study, we have investigated the
confinement-deconfinement nature in the effective Polyakov-line model
via the persistent homology analysis on the rectangular lattice.  By
using the persistent homology analysis, we can investigate the behavior
of the multiscale spatial distribution of data inside the center
clusters.  To compute the persistent homology, we divide the complex
Polyakov-line plane into three domains, ${\cal Z}_0$, ${\cal Z}_1$ and
${\cal Z}_2$, and then the lattice data are mapped on the plane.  We
then prepare the ${\cal Z}_0$, ${\cal Z}_1$ and ${\cal Z}_2$ lists to
storage the corresponding data.  Our results are the following:
\begin{enumerate}
 \item In the confined phase, the data are uniformly distributed on
       each list and thus we have similar persistent diagram for each
       list.
       This means that each list has same topological data
       structure.
       Typical shapes appearing in the data space are analyzed;
       actual shapes are depicted in Fig.~\ref{fig:ph_comic}.  Comparing
	   with the results of the random configuration with the occupation
	   rate 33\%, we find that some holes with the death time $3.5
	   \lesssim d \lesssim 5$, which do not appear in the random model,
	   exist in the confined phase for the EPL model, and the one of the
	   dominant structures is different from that of the random
	   configuration.  These results may indicate the existence of the
	   nontrivial topological structures.
 \item In the deconfined phase, the data are dense in the ${\cal Z}_0$
	   domain, but they become sparse in the ${\cal Z}_1$ and
       ${\cal Z}_2$ domains.
	   Due to this property,
	   the ${\cal Z}_0$ and ${\cal Z}_{1,2}$ lists have
       different topological structures in the data space.
       In the case of the ${\cal Z}_0$ list, the smallest $1\times 1\times 1$
       cube and many small structures appear.
       In comparison, the ${\cal Z}_{1,2}$ lists have the structures
       which have late birth time and short life time.  Comparing the
	   result of the ${\cal Z}_0$ domain with the results of the random
	   configuration with the occupation rate 98\%, the persistent
	   diagram and the dominant structures are almost identical.  The
	   distribution of data points of the ${\cal Z}_0$ domain in the
	   deconfined phase are qualitatively described by the randomly
	   distributed model.
 \item To clearly show the phase transition, we consider the
       configuration averaged ratio of the birth and death time.
       This quantity shows the quite different behavior in the confined
       and the deconfined phases.
       In particular, the ratios of the ${\cal Z}_0$ and ${\cal Z}_{1,2}$ lists
       depart each other when the deconfined properties appear in the system.
       The difference comes from the structural change of the
       data by the confinement-deconfinement transition.  Furthermore,
	   this quantity in the EPL model differs from that of the random
	   configuration with the same occupation rate.  This difference may
	   indicate the existence of
	   the more complex multiscale topological structure
	   of the domain.
 \item By considering the configuration averaged ratio of the birth and
	   death times of holes, we can construct the nonlocal order-parameter of the
       confinement-deconfinement transition from the topological viewpoint of
       the data space.  The statistical error of this quantity is much
	   smaller than that of the expectation value of Polyakov-line.
	   This fact may indicate that the structures in the configurations
	   have the similar spatial distribution.
	   Therefore, by using this quantity,
	   we can characterize the phases with a fewer configurations than
	   that of Polyakov-line.
\end{enumerate}

We can consider the following interesting and important future works:
 \begin{enumerate}
  \item In Refs.~\cite{Gattringer:2010ms,Borsanyi:2010cw,Endrodi:2014yaa,Schafer:2015wja},
		the center clusters are investigated by isolating the lattice
		with the narrower range of the Polyakov-line phases than ours.
		It is interesting to combine this idea with our persistent
		homology method.  In particular, to investigate more tiny
		structures in the deconfined phase, it is interesting to isolate
		the lattice into three domains according to the prescription
		presented in Ref.~\cite{Gattringer:2010ms}, and we analyze the
		multiscale spatial structure for each domain by using persistent
		homology.
  \item In this study, we do not investigate so much near the phase
		transition point.  The persistent homology analyses near the
		phase transition point should be necessary.  In particular, the
		dependence of the persistent homology against the order of the
		phase transition is interesting.
  \item The persistent homology is based on the spatial
		topology with the center clustering in this work. In this sense,
		this determination can treat not only the global feature of the
		spontaneous ${\mathbb Z}_3$ symmetry breaking but also the spatial
		distribution itself (see Figs.~\ref{fig:map_data_k0120} and
		\ref{fig:map_data_k0150}), while the ordinary determination of
		the confinement-deconfinement nature of QCD is done by using the
		averaged value of the Polyakov-line (see
		Fig.~\ref{fig:ploop_kappa-dep}). In some cases, the present
		determination and the ordinary determination based on
		the Polyakov-line should provide the same feature of QCD, but the
		present determination may provide more information for the
		confinement-deconfinement nature of QCD if there is the
		topological structure in the spatial structure. Of course, we
		need more detailed studies on it, particularly in QCD with
		dynamical quarks, and hence it is our future work to be done.
  \item Studying the spatial structures of the center vortex is also an
		interesting subject.  It is considered that the center vortex
		goes through the loop constructed by the center projected link
		variables with nontrivial values, and that it is related
		strongly with the mechanism of the confinement-deconfinement
		nature
		\cite{greensite2010introduction,Greensite:2003bk,Engelhardt:2004pf}.
		By using the persistent homology, we may obtain the hints to
		describe the mechanism of the quark confinement from the
		viewpoint of the multiscale spatial structures of center
		vortices.
 \end{enumerate}
These results will be shown in elsewhere.

\section*{Acknowledgments}
 The authors are thankful to Masahiro Ishii, Akihisa Miyahara and Akira
 Ohno for fruitful discussions.  This study is supported in part by the
 Grants-in-Aid for Scientific Research from JSPS (No.~18K03618 and
 No.~17K05446).

\bibliography{./ref}

\end{document}